\documentclass[
aip, 
reprint,
citeautoscript,
altaffilsymbol,
floatfix]{revtex4-1}
\usepackage{graphicx}
\usepackage[colorlinks=true, linkcolor=green, citecolor=blue]{hyperref}
\usepackage{xcolor}
\usepackage[T1]{fontenc}
\usepackage{amsmath,amsfonts}
\usepackage{array}
\usepackage{tabularx}
\usepackage{ragged2e}

\usepackage{setspace}
\usepackage{float}
\usepackage{tikz}
\usepackage{dirtytalk}
\usetikzlibrary{shapes.geometric, positioning, arrows.meta}
\usepackage{adjustbox}\usepackage[final]{pdfpages}
\usepackage{pgffor}
\usepackage{booktabs}
\usepackage{adjustbox,makecell,booktabs}
\usepackage{geometry}
\usepackage{siunitx}
\usepackage[utf8]{inputenc}
\usepackage{lipsum}
\newcommand{\vaspkw}[1]{  \begingroup
  \setlength{\fboxsep}{1pt}  \colorbox{gray!15}{\strut\texttt{#1}}  \endgroup
}\renewcommand{\selectlanguage}[1]{}

\newcolumntype{Y}{>{\Centering\arraybackslash}X}
\newcolumntype{P}[1]{>{\Centering\arraybackslash}p{#1}}
\newcolumntype{C}[1]{>{\Centering\arraybackslash}m{#1}}
\newcolumntype{M}[1]{>{\RaggedRight\arraybackslash}m{#1}}

\makeatletter
\def\fps@figure{t}
\AtBeginDocument{\let\LS@rot\@undefined}
\makeatother
\makeatletter
\let\@fnsymbol\@fnsymbol@latex
\@booleanfalse\altaffilletter@sw
\def\frontmatter@makefnmark{ \@textsuperscript{\normalfont\@thefnmark}}
\makeatother

\begin{document}

\title{Dyna-Mat: End-to-end benchmarking of foundation machine learning interatomic potentials in finite-temperature ensembles}

\author{Miko{\l}aj J. Gawkowski}
\affiliation{Department of Physics and Astronomy, University College London, 7-19 Gordon St, London WC1H 0AH, UK
}
\affiliation{Thomas Young Centre and London Centre for Nanotechnology,
9 Gordon St, London WC1H 0AH, UK
}

\author{Nongnuch Artrith}
\altaffiliation{Listed in alphabetical order.}
\affiliation{
Debye Institute for Nanomaterials Science, Utrecht University, 3508 TA Utrecht, the
Netherlands
}

\author{Silvia Bonfanti}
\altaffiliation{Listed in alphabetical order.}
\affiliation{NOMATEN Centre of Excellence, National Center for Nuclear Research, ul. A. So\l{}tana 7, 05-400 Swierk/Otwock, Poland.}

\author{Abhijeet Sadashiv Gangan}
\altaffiliation{Listed in alphabetical order.}
\affiliation{Periodic Labs, Menlo Park, CA, USA}

\author{Hendrik H. Heenen}
\altaffiliation{Listed in alphabetical order.}
\affiliation{Fritz-Haber-Institut der Max-Planck-Gesellschaft, Faradayweg 4-6, D-14195 Berlin, Germany}

\author{Joseph Kioseoglou}
\altaffiliation{Listed in alphabetical order.}
\affiliation{
Physics Department, Aristotle University of Thessaloniki, GR-54124 Thessaloniki, Greece
}
\affiliation{
Center for Interdisciplinary Research and Innovation, Aristotle University of Thessaloniki, GR-57001 Thessaloniki, Greece
}

\author{Ivor Lončarić}
\altaffiliation{Listed in alphabetical order.}
\affiliation{Ru{\dj}er Bo\v{s}kovi'c Institute, Bijeni\v{c}ka cesta 54, 10000 Zagreb, Croatia}

\author{Hemanadhan Myneni}
\altaffiliation{Listed in alphabetical order.}
\affiliation{The Faculty of Industrial Engineering, Mechanical Engineering, and Computer Science, University of Iceland, Reykjavík, Iceland}

\author{Janosh Riebesell}
\altaffiliation{Listed in alphabetical order.}
\affiliation{Periodic Labs, Menlo Park, CA, USA}

\author{Mariana Rossi}
\altaffiliation{Listed in alphabetical order.}
\affiliation{
MPI for the Structure and Dynamics of Matter, Luruper Chaussee 148, 22761 Hamburg, Germany
}
\affiliation{
Yusuf Hamied Department of Chemistry, Lensfield Road, CB2 1EW Cambridge, UK
}

\author{Matthias Rupp}
\altaffiliation{Listed in alphabetical order.}
\affiliation{Luxembourg Institute of Science and Technology (LIST), 5 Avenue des Hauts-Fourneaux, L-4362 Esch-sur-Alzette, Luxembourg
}

\author{Jonathan Schmidt}
\altaffiliation{Listed in alphabetical order.}
\affiliation{Laboratory of Computational Science and Modeling, Institut des Mat'eriaux, 'Ecole Polytechnique F'ed'erale de Lausanne, 1015 Lausanne, Switzerland}

\author{Shubham Sharma}
\altaffiliation{Listed in alphabetical order.}
\affiliation{
MPI for the Structure and Dynamics of Matter, Luruper Chaussee 148, 22761 Hamburg, Germany
}

\author{Benjamin X. Shi}
\altaffiliation{Listed in alphabetical order.}
\affiliation{Initiative for Computational Catalysis, Flatiron Institute, New York, NY 10010, USA}

\author{Antoni Wadowski}
\altaffiliation{Listed in alphabetical order.}
\affiliation{NOMATEN Centre of Excellence, National Center for Nuclear Research, ul. A. So\l{}tana 7, 05-400 Swierk/Otwock, Poland.}
\affiliation{Faculty of Materials Science and Engineering, Warsaw University of Technology, Wo\l{}oska 141, 02-507 Warsaw, Poland.}

\author{Lukas H\"{o}rmann}
\email{lukas.hoermann@univie.ac.at}
\affiliation{Faculty of Physics, University of Vienna, Vienna, 1090, Austria}
\affiliation{Department of Chemistry, University of Warwick, Coventry, CV4 7AL, United Kingdom}

\author{Venkat Kapil}\email{v.kapil@ucl.ac.uk}
\affiliation{Department of Physics and Astronomy, University College London, 7-19 Gordon St, London WC1H 0AH, UK
}
\affiliation{Thomas Young Centre and London Centre for Nanotechnology,
9 Gordon St, London WC1H 0AH, UK
}

\newpage
	
\begin{abstract}
Foundation machine learning interatomic potentials (MLIPs) are increasingly being used as drop-in replacements for first-principles calculations -- enabling simulations of materials at length and time scales that were previously inaccessible. 
However, due to lack of ground truth data, their accuracy on structural and dynamical observables in finite thermodynamic ensembles is yet to be established.
Here, we introduce \textsf{Dyna-Mat-v1.0}, a benchmark dataset of condensed-phase first-principles molecular dynamics trajectories designed to test foundation MLIPs at realistic finite-temperature conditions. 
Using this dataset, we evaluate 15 foundation MLIPs across four model tiers by comparing both single-point energy and force errors on first-principles configurations and observables generated from MLIP-driven trajectories. 
We find that "on average" models with lower single-point force errors also yield lower errors for structural and dynamical observables. 
However, there are individual systems for which low force errors lead to qualitative failures in the predicted structure. 
Pressure remains poorly described across most models, pointing to limitations in the density functional theory stress labels available in current large-scale training datasets. 
Finally, we construct an accuracy-cost Pareto frontier to identify the best trade-offs for molecular dynamics with foundation MLIPs, finding that the latest generation of cross-trained models is (close to) Pareto-optimal according to the accuracy metrics considered here. 
Overall, \textsf{Dyna-Mat-v1.0} shows that end-to-end finite-temperature validation is essential for quantifying the predictive behaviour of foundation MLIPs, and provides a simple, scalable route for assessing them beyond static and harmonic benchmarks relevant to materials design.
\end{abstract}
	
\maketitle

\newpage

\section{Introduction}

\noindent Machine learning interatomic potentials (MLIPs)~\cite{behler_four_2021,deringer_gaussian_2021,jacobs_practical_2025} -- trained on datasets comprising energies and gradient labels generated by density functional theory (DFT) --  are rapidly transforming atomistic materials modelling~\cite{jacobs_practical_2025, yuan_foundation_2026, creed_six_2026}. 
A recent development is the emergence of so-called foundation MLIPs, which are broadly trained models that can either be fine-tuned efficiently for bespoke applications~\cite{kaur2024dataefficientfinetuningfoundationalmodels,novelli_fast_2025,radova_fine-tuning_2025, hanseroth_fine-tuning_2026, tompa_fine-tuning_2026} or used zero-shot for materials screening~\cite{gnome, li_probing_2025, inizan_system_2025}. 
These models are attractive for materials design, where rapid exploration of chemical, structural, and thermodynamic design spaces is often limited by the computational cost of first-principles simulations. 
However, predicting \textit{a priori} whether a foundation MLIP reproduces DFT-quality results is challenging, because the reliability of a model is sensitive to the architecture, the size, diversity, and quality of the training set, and the target observable. \\

\noindent In this context, community benchmarks have become central to assessing the reliability of foundation MLIPs.  
Matbench-discovery by Riebesell et al.~\cite{riebesell_framework_2025}, amongst others~\cite{chiang_mlip_2025,choudhary_jarvis-leaderboard_2024, peng_lambench_2026}, has played a pivotal role in steering their development. 
These efforts now cover structural relaxation and phase-stability classification~\cite{riebesell_framework_2025}, static energies and forces, elastic constants, defects, surfaces, interfaces, phonons, thermal conductivity~\cite{pota2024thermal}, catalysis~\cite{moon_catbench_2025}, high-pressure behaviour~\cite{loew2026universal}, and porous materials~\cite{kras_mofsimbench_2025}. 
Several studies have identified clear failure modes, with Focassio et al.~\cite{focassio_performance_2025} reporting large errors in out-of-distribution surface energetics, Deng et al.~\cite{deng_systematic_2025} finding systematic softening of barriers, phonon modes, and formation energies, and Loew et al.~\cite{loew2026universal} showing that transferability deteriorates under pressure. 
The CHIPS-FF benchmark of Wines and Choudhary~\cite{wines_chips-ff_2025} further showed that newer models trained on larger and more diverse datasets, including the union of \textsf{MPTrj}~\cite{chgnet} and \textsf{Alexandria}~\cite{dataset-alexandria}, \textsf{MatterSim}~\cite{mattersim-v1}, and \textsf{OMat24}~\cite{dataset-omat24}, improve surface and defect formation energies, while interfacial work of adhesion remains poorly predicted across all tested models. 
These benchmarks have motivated the creation of new datasets that incorporate such diversity, leading to improved MLIPs~\cite{malosso_high-quality_2026, kaplan_foundational_2025}. \\

\noindent A common denominator associated with current benchmarks is their reliance on single-point energies and forces or (corrections built on) the harmonic approximation. 
These benchmarks therefore do not directly assess how well foundation MLIPs reproduce structural and dynamical properties of materials at finite thermodynamic conditions, such as via molecular dynamics (MD).
Existing studies have partly addressed the reliability of MD trajectories generated by foundation MLIPs, either by evaluating force errors against DFT labels on the sampled configurations~\cite{fragapane_lips_2026}, or by comparing observables obtained from these trajectories and those generated by system-specific MLIPs trained from scratch and fine-tuned, with agreement between the models serving as a consistency check~\cite{pet-oam}.
A limitation of these validations is that the ensemble from which configurations are sampled for probing errors or for fine-tuning may be biased due to errors in the MLIPs against the DFT reference.  \\

\begin{figure*}[t]
    \centering
    \includegraphics[width=\linewidth]{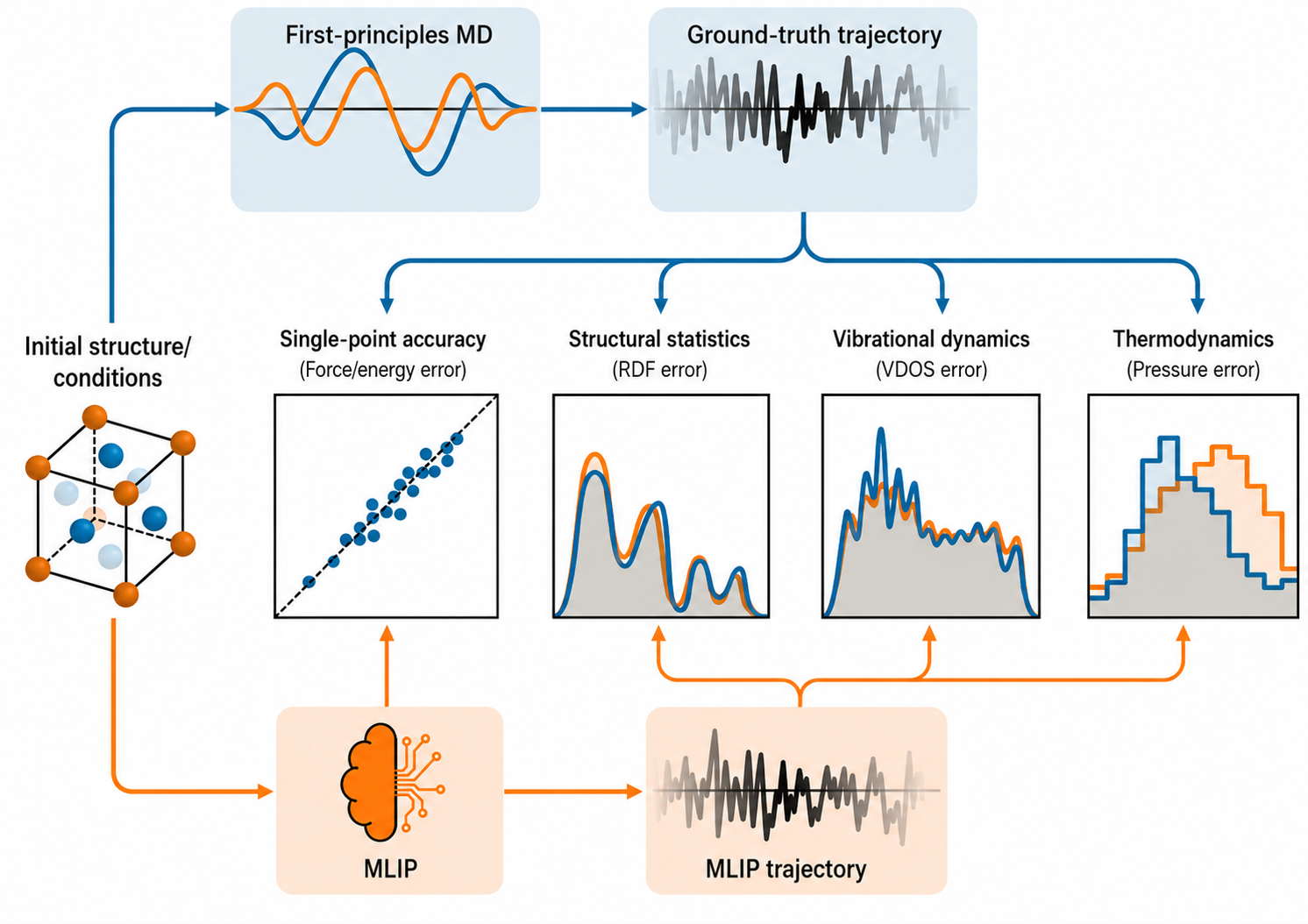}
    \caption{Schematic of the dataset generation and benchmarking workflow. Starting from a consistent set of initial structures and simulation settings, first-principles and MLIP molecular dynamics simulations are performed. The reference first-principles trajectory is used to quantify single-point energy and force accuracy, as well as structural and dynamical properties such as radial distribution functions, vibrational density of states, and pressure distributions.}
    \label{fig:toc}
\end{figure*}

\noindent In this work, we present a benchmark dataset comprising reference finite-temperature ensembles known as \textsf{Dyna-Mat-v1.0} -- generated from first-principles MD for a diverse set of periodic materials -- and an assessment of foundation MLIPs on zero-shot description of structural and dynamical properties at finite temperature. 
Although generating these trajectories has been a computationally demanding endeavour, we hope that its cost will be amortised by providing a common reference against which many foundation MLIPs can be evaluated, thereby reducing the volume of bespoke validations. 
We consider 15 models spanning four generations of training-set size and model complexity, and evaluate their ability to reproduce observables, including the radial distribution functions, pressures, and vibrational densities of states. 
By comparing these finite-temperature errors with standard energy- and force-based metrics, as well as with figures of merit from existing community benchmarks, we assess which widely used metrics are predictive of finite-temperature accuracy and where current foundation MLIPs remain limited.

\section{Results}

\subsection{The \textsf{Dyna-Mat-v1.0} dataset}

\noindent As an outcome of the efforts of working groups 2 and 4 of the Data-driven Applications towards the Engineering of Functional Materials COST Action, we present the \textsf{Dyna-Mat-v1.0} dataset. 
It consists of 17 first-principles MD trajectories in the $\mathrm{NVT}$ ensemble across six classes of materials: pure metals, transition metal dichalcogenides, metal alloys, perovskites, and molecular crystals, at temperatures in the range 293-1500\,K and for simulation times of up to $\sim$21\,ps. \\

\begin{figure*}[t]
    \centering
    \includegraphics[width=1\linewidth]{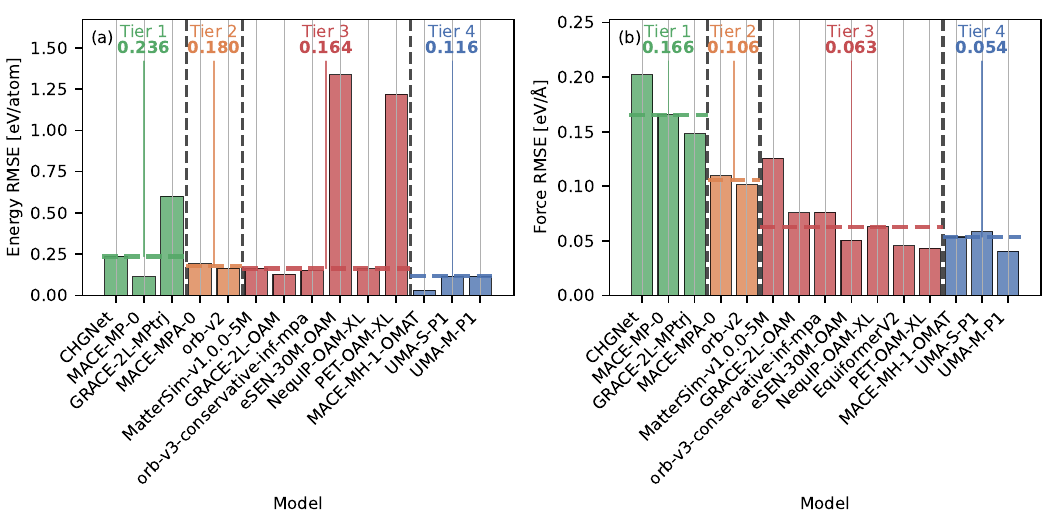}
    \caption{(a) Mean energy RMSE and (b) mean force RMSE for each model averaged over all 17 systems. Within each tier, the models are shown according to the increasing number of model parameters. \texttt{EquiformerV2} is excluded from the energy RMSE plot. Horizontal dashed lines indicate the median mean energy or force RMSE for each model tier.
        }
    \label{fig:plot-energy-force-rmses}
\end{figure*}

\noindent To ensure consistency, all simulations with the exception of molecular crystals, were carried out using a common set of converged DFT settings using the VASP software~\cite{hafner_ab-initio_2008}.
The simulations for molecular crystals were performed using the FHI-aims software~\cite{blum_ab_2009} and i-PI~\cite{litman_i-pi_2024}, but at a consistent level of DFT convergence. 
Starting from the MPRelax settings in Pymatgen~\cite{ong_python_2013} used to generate the \textsf{MPTrj} dataset, the plane-wave cutoff and $k$-point grid density were increased until the total energies and forces were converged to 2.0 meV/atom and 50 meV/\AA{} (these tests were carried out for a subset of systems).
A common command-line interface based on \texttt{atomate2}~\cite{ganose_atomate2_2025} and \texttt{ASE}~\cite{ASE} was used to reproducibly perform these simulations, with the exception of molecular crystals, for which simulations were performed using \texttt{FHI-aims} and \texttt{i-PI} v3.1.3~\cite{litman_i-pi_2024}. 
The starting configurations were obtained by fixed-cell structure optimisation using experimental lattice parameters. 
More details of the computational settings are provided in the Methods section (Sec.~\ref{sec:methods}).

\subsection{Single-point energy and force errors for foundation MLIPs}

\noindent With access to the \textsf{Dyna-Mat-v1.0} dataset, we begin by benchmarking a set of 15 foundation MLIPs on two simple metrics. 
We consider the energy and force root mean square errors (RMSEs), obtained using Eq.~(\ref{eq:energy-rmse}) and Eq.~(\ref{eq:force-rmse}), of the models evaluated over the full trajectory as single-point calculations.
The models are classified into four Tiers based on the size or diversity of the dataset(s) they were (pre)trained on. 
\begin{enumerate}
    \item Tier 1 includes models trained on the \textsf{MPTrj} dataset: \texttt{CHGNet}~\cite{chgnet}, \texttt{MACE-MP-0}~\cite{mace-mp-0} and \texttt{GRACE-2L-MPtrj}~\cite{grace-mptrj}.
    \item Tier 2 includes models trained on combined \textsf{MPTrj}$+$\textsf{Alexandria} (or the subsampled \textsf{Alexandria}): \texttt{MACE-MPA-0}~\cite{mace-mp-0}, which was trained directly, and \texttt{orb-v2}~\cite{orb-v2}, which was fine-tuned on the said datasets after diffusion pretraining. 
    \item Tier~3 includes models trained on datasets larger and more diverse than those used for Tier~2: \texttt{MatterSim-v1.0.0-5M}~\cite{mattersim-v1}, \texttt{GRACE-2L-OAM}~\cite{grace-oam}, \texttt{orb-v3-conservative-inf-mpa}~\cite{orb-v3}, \texttt{eSEN-30M-OAM}~\cite{esen-oam}, \texttt{NequIP-OAM-XL}~\cite{nequip-oam}, \texttt{EquiformerV2}~\cite{liao_equiformerv2_2024}, and \texttt{PET-OAM-XL}~\cite{pet-oam}. Most of these models are associated with \textsf{OAM}-type training data, which combines \textsf{OMat24} pretraining with fine-tuning on \textsf{MPTrj}$+$\textsf{Alexandria} or a subsampled variant thereof. \texttt{orb-v3-conservative-inf-mpa} follows a related but distinct protocol, starting from diffusion pretraining, followed by fine-tuning on a subset of \textsf{OMat24} and then on \textsf{MPTrj}$+$\textsf{Alexandria}. \texttt{MatterSim-v1.0.0-5M} is trained on the \textsf{MatterSim} dataset, while \texttt{EquiformerV2} is pretrained on \textsf{OMat24} and fine-tuned on \textsf{MPTrj}.
    \item Tier 4 consists of multi-head or multi-task models trained across multiple reference datasets and electronic-structure settings: \texttt{MACE-MH-1-OMAT}~\cite{mace-mh-models}, which is pretrained on \textsf{OMat24} and multi-head fine-tuned on datasets including \textsf{OMat24 Replay} (10\% of \textsf{OMat24}), \textsf{OMol-1\%}, \textsf{OC20}, \textsf{SPICE-1}, \textsf{RGD1}, \textsf{MPTrj}, and \textsf{MatPES}; and \texttt{UMA-S-P1} and \texttt{UMA-M-P1}~\cite{uma-models}, which are trained on the combined \textsf{OMat24}, \textsf{OMol25}, \textsf{OC20++}, \textsf{OMC25}, and \textsf{ODAC25} datasets. For both \texttt{UMA} models, we use the \texttt{p1} checkpoints.
\end{enumerate}
It is worth noting that the pseudopotentials used in \textsf{MPTrj} and \textsf{Alexandria} are inconsistent with those used in \textsf{OMat24}, which further motivates separating Tier~4 models from Tier~3. \\

\noindent In Fig.~\ref{fig:plot-energy-force-rmses}(a), we report the mean energy RMSE evaluated on configurations sampled from the \textsf{Dyna-Mat-v1.0} trajectories.
These errors are computed from single-point energy and force predictions on each MD snapshot, and therefore do not yet assess finite-temperature observables.
Given molecular crystal simulations were performed with FHI-aims as opposed to VASP, we estimate energy RMSEs by referencing both the MLIP and DFT total energies to their respective isolated-atom limits by subtracting the stoichiometric sum of the isolated-atom energies from each crystal configuration.
Isolated-atom energies could not be computed with \texttt{EquiformerV2} because a single atom produces an empty neighbour graph, which triggers a runtime failure. Therefore, we omit \texttt{EquiformerV2} from Fig.~\ref{fig:plot-energy-force-rmses}(a).
To disentangle, as far as possible, the effects of model architecture and training-set size, we discuss the median error within each tier separately.
The median energy RMSE decreases from 0.236~eV/atom for Tier 1 to 0.180~eV/atom for Tier 2.
Tier 3 exhibits some outliers with high energy errors (discussed later), which result in a marginal reduction in the RMSE to 0.164~eV/atom for Tier 3 compared to Tier 2. 
Finally, Tier 4 has the lowest median RMSE of 0.116~eV/atom. \\

\noindent To contextualise these values, we further break down the errors by material class.
As shown in Fig.~\ref{fig:plot-SI-energy-rmse}, the largest errors are observed for molecular crystals, for which the highest energy RMSE is obtained with the Tier-3 model \texttt{eSEN-30M-OAM}, and the lowest with the Tier-4 model \texttt{MACE-MH-1-OMAT}.
There are two possible explanations for the large errors observed for molecular crystals. 
First, they may arise from the presence of high-frequency intramolecular modes and, second, from the limited representation of molecular crystals in the training datasets.
For perovskites and TMDs, models trained on more extensive datasets show large reductions in energy errors, whereas for metal alloys the corresponding improvements are more modest, with the lowest median errors obtained by Tier-4 models. 
For pure metals, the median tier-wise energy RMSE does not improve from Tier 1 to Tier 3, although \texttt{EquiformerV2} is a notable outlier, yielding errors nearly an order of magnitude lower than the other models across Tier 1-3. 
All Tier-4 models perform substantially better for pure metals, with energy RMSEs more than an order of magnitude lower than those of the lower tiers. 
Again, it is worth highlighting that the pseudopotentials used in our reference calculations are consistent with those used by Tier~4 models, which may contribute to their lower errors. \\

\noindent Because energy RMSE can be dominated by absolute energy offsets (which may arise from the lack of a baseline atomic energy in some models or the difference in the DFT settings across datasets) rather than errors in the sampled statistical ensemble, we refrain from interpreting this metric in isolation.
We instead turn to force RMSEs, which drive sampling errors within MD.
In Fig.~\ref{fig:plot-energy-force-rmses}(b), we observe a systematic improvement in median force RMSE across newer generations of foundation MLIPs.
The median force RMSE decreases from 0.166~eV/\AA{} for Tier~1 to 0.106~eV/\AA{} for Tier~2, further decreases to 0.063~eV/\AA{} for Tier~3, and reaches its lowest value of 0.054~eV/\AA{} for Tier~4. \\

\noindent These median force RMSEs suggest that improvements in force prediction across the tiers are more systematic than those observed for energy RMSEs.
Inspecting Fig.~\ref{fig:plot-SI-force-rmse}, we observe that molecular crystals exhibit the highest median force errors, which could be attributed either to higher-frequency intramolecular modes compared to inorganic crystals or that these crystals are predominantly stabilised by long-range van-der-Waals interactions that are not fully captured by local MLIPs, while pure metals and perovskites yield the lowest median force RMSEs.
To gain further insight, we also analyse the energy and force RMSEs after excluding the molecular-crystal trajectories, as shown in Fig.~\ref{fig:plot-SI-energy-force-rmses-no-molecular-crystals} of the SI. That allows us to include \texttt{EquiformerV2} in the energy RMSE plot.
We observe that the energy errors decrease from Tier~3 to Tier~4, whereas the force errors remain approximately saturated across these tiers.
This suggests that, after excluding molecular crystals, the apparent improvement in energy RMSE primarily reflects a reduction in constant energy offsets rather than a corresponding improvement in local force accuracy.
For molecular crystals, by contrast, the improvement appears to involve not only energy offsets but also fluctuations of the potential-energy surface. \\

\noindent The tier-wise reduction in force RMSE can be indicative of improved accuracy of MD ensembles, and therefore it sets the stage for evaluating whether these lower errors translate into more accurate structural, thermodynamic, and dynamical properties.

\begin{figure*}[t]
    \centering
    \includegraphics[width=1\linewidth]{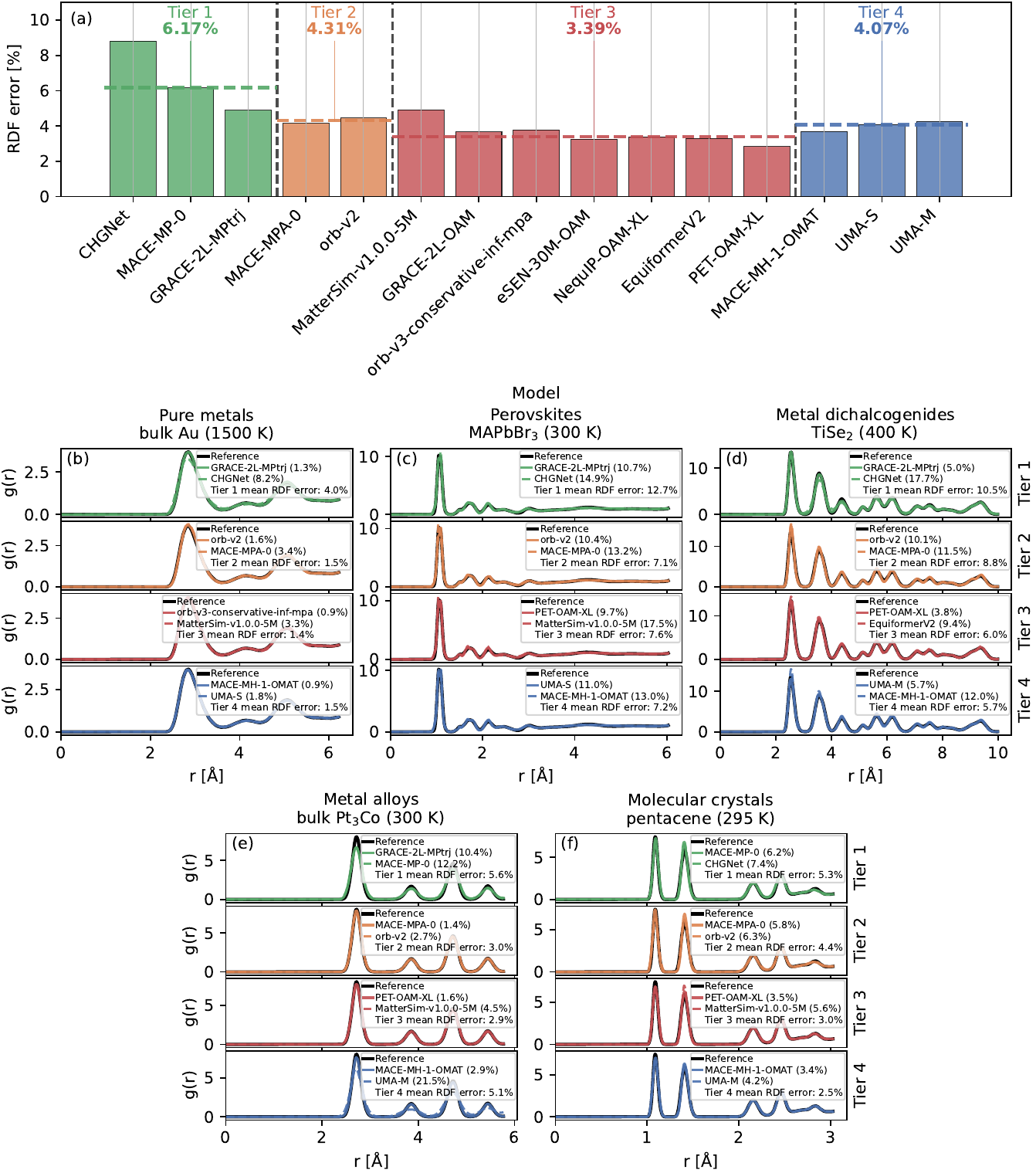}
    \caption{
    Radial distribution function $g(r)$ panel.
    (a): RDF error values for the models benchmarked in this work. The models are grouped into four tiers as defined in the main text. The horizontal dashed lines indicate the tier-wise median RDF error values.
    (b--f):  RDF is given for the worst-performing system from each material class, and in each case, the reference RDF is plotted along with the best (solid lines) and worst (dashed lines) performing models on the given system for each system tier. The material classes are: b) pure metals, c) perovskites, d) metal dichalcogenides, e) metal alloys, and f) molecular crystals.}
    \label{fig:plot-rdf-panel}
\end{figure*}

\subsection{Observable errors for foundation MLIPs}

\noindent We next perform MD simulations with each foundation MLIP using the same initial structures, temperature, and simulation settings as the reference first-principles trajectories, including thermostat type and relaxation constant. 
The generated trajectories are post-processed to estimate observables which are then compared with their first-principles counterparts.

\subsubsection{Radial distribution functions}

\noindent We first consider the radial distribution function (RDF), which we estimate for all atom types relative to all atoms (see Eq.~(\ref{eq:rdf-def})). 
This quantity is a comprehensive probe of global relative positions and is applicable to generic systems.
To facilitate comparisons of RDFs generated from MLIP trajectories among one another and with the \textsf{Dyna-Mat-v1.0} reference, we define a single-valued percentage error metric for the RDF (see Eq.~(\ref{eq:rdf-error})).
The RDF error is designed to yield 0~\% for a perfect match and 100~\% for an RDF generated by an ideal gas. \\

\noindent Fig.~\ref{fig:plot-rdf-panel}(a) summarises the RDF error, averaged over all trajectories, for all models and classifies them by tiers.
Consistent with the force RMSEs evaluated on the \textsf{Dyna-Mat-v1.0} trajectories, median RDF errors generally decrease as one moves from lower to higher tiers, with the largest reduction observed between Tiers~1 and 2 and diminishing returns beyond Tier~2.
In fact, the median errors across Tiers 2-4 are similar, with Tier 3  \texttt{PET-OAM-XL} displaying the lowest errors of all models.  \\

\noindent To understand if the errors are dominated by any particular material class, we also study the reference and predicted RDFs per material class. 
In Fig.~\ref{fig:plot-rdf-panel}(b--f), we show the predicted vs reference RDFs for the worst-performing system from each material class. 
In each case, we report the best- and worst-performing MLIPs for each system, by providing the RDF error in the legend as defined in Eq. (\ref{eq:rdf-error}).
Additionally, for each material class, we report the tier-wise mean RDF error.
The lowest mean RDF error is observed for Tiers 3 and 4 on pure metals. For bulk Au, the best-performing models are \texttt{orb-v3-conservative-inf-mpa} and \texttt{MACE-MH-1-OMAT}, both with an RDF error of only 0.9\%.
Across most other classes, the model tiers achieve sub-10\% mean RDF errors. 
The only exceptions are Tier 1 for perovskites and TMDs, for which the mean RDF errors are 12.7\% and 10.5\%, respectively.
The case of bulk Pt$_3$Co is interesting as Tier 4 has a lower mean RDF error than Tier 1. 
At the same time, however, if we compare the worst-performing models within their respective Tiers, \texttt{UMA-M-P1} yields a high RDF error of 21.5\%, compared with 12.2\% for \texttt{MACE-MP-0}, indicating that \texttt{UMA-M-P1} is an outlier for this system.
A closer look at the RDF suggests a softening of the structure compared to the reference.  \\

\noindent Benchmarking against the RDF is challenging because the finite-temperature structures are probed at temperatures up to 1500~K. H
However, the benchmark complexity is reduced by performing the MLIP trajectories at a fixed unit cell matching the reference first-principles simulations, which motivates the next analysis. 

\subsubsection{Pressures}
\begin{figure*}[t]
    \centering
    \includegraphics[width=1\linewidth]{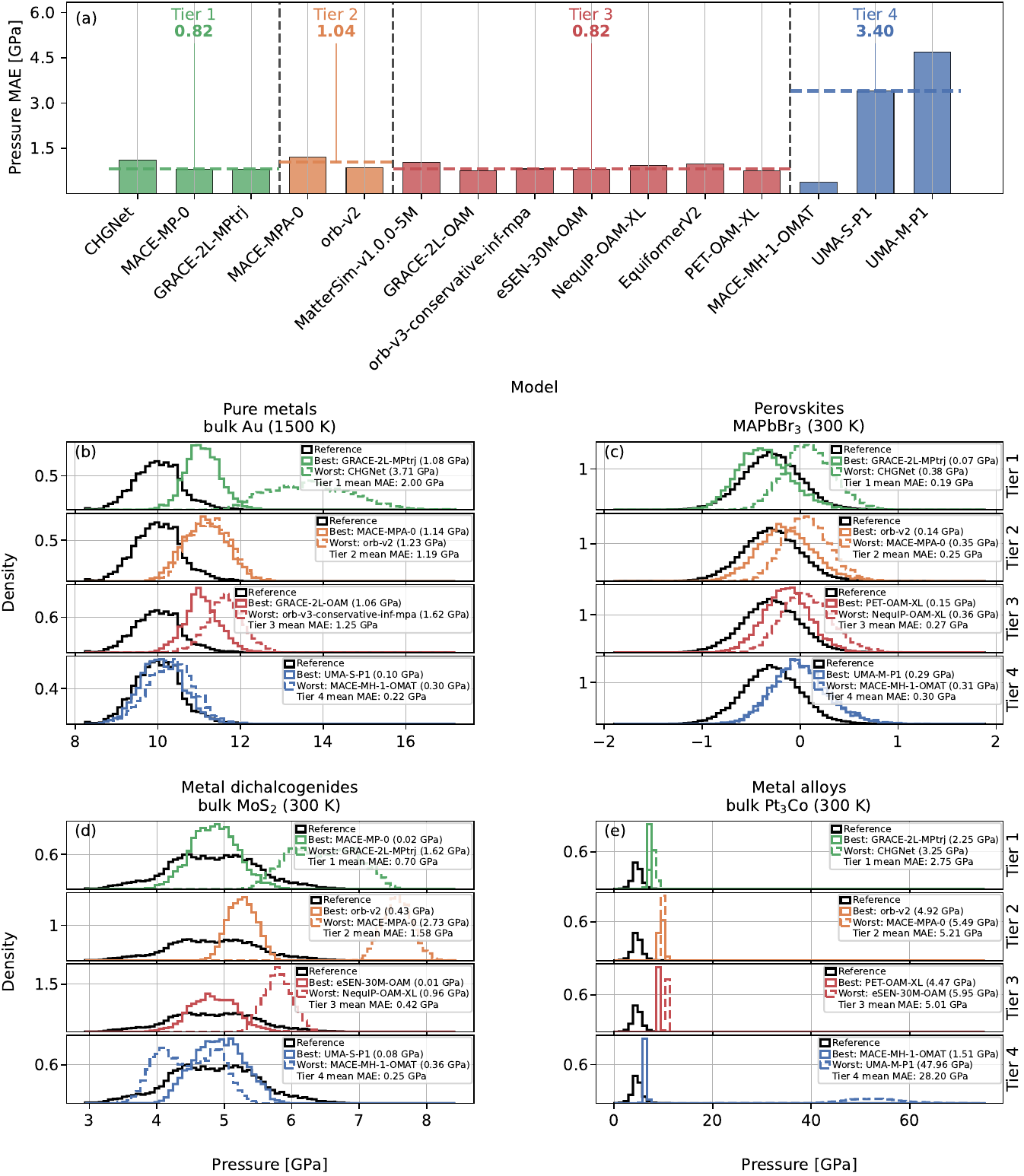}
    \caption{
    Unit-cell pressure distribution panel.
    (a): pressure MAE for the MLIPs grouped in four tiers. The MAE is taken with reference to the first-principles MD simulations. Molecular crystals are not included. Horizontal dashed lines indicate the mean pressure MAE for each model tier.
    (b-e): pressure histograms for the worst-performing system from each material class. Each subplot shows the trajectory pressures for the best- and worst-performing models from each model tier, as well as the reference one. The material classes considered here are: b) pure metals, c) perovskites, d) metal dichalcogenides, and e) metal alloys.}
    \label{fig:plot-pressure-panel}
\end{figure*}
\noindent Pressure provides a test complementary to the RDFs estimated in a fixed-cell ensemble because its mean and fluctuations probe the tendency of the model to drive lattice expansion and contraction in the $\mathrm{NPT}$ ensemble~\cite{tuckerman_statistical_2010}.
We therefore estimate the unit-cell pressure distribution for each trajectory.
The reference pressure is read off from the reference simulation output files, while the pressure associated with the MLIPs is consistently calculated from the trace of the virial tensor.
For each system, pressure error is obtained using Eq.~(\ref{eq:pressure-mae-error-per-system}) and averaged for each MLIP using Eq.~(\ref{eq:pressure-mae-error-per-model}). We report the pressure error in GPa, and we also visualise the full pressure distributions as histograms. \\

\noindent The pressure mean absolute error (MAE), averaged over all systems, is summarised in Fig.~\ref{fig:plot-pressure-panel}(a) for all models classified by their tiers. 
In contrast to the RDF errors, we do not observe a systematic improvement in the pressure MAE with model tier. 
The median pressure MAE for Tier~1 is 0.82~GPa, and increases by 0.22~GPa for Tier~2.
Tier~3 attains the same median pressure MAE as Tier~1, indicating no systematic improvement in pressure prediction with increasing model size or training-set scale.
Tier~4 exhibits the highest median pressure MAE, 3.40~GPa, mainly due to the large pressure errors of the \texttt{UMA} models.
Nevertheless, \texttt{MACE-MH-1-OMAT} is a notable exception, achieving the lowest pressure MAE overall.
In Fig.~\ref{fig:plot-SI-pressure-violin-plot}, we additionally compare the predicted and reference pressure distributions for each model via split violins. 
\\

\begin{figure*}[t]
    \centering
    \includegraphics[width=1\linewidth]{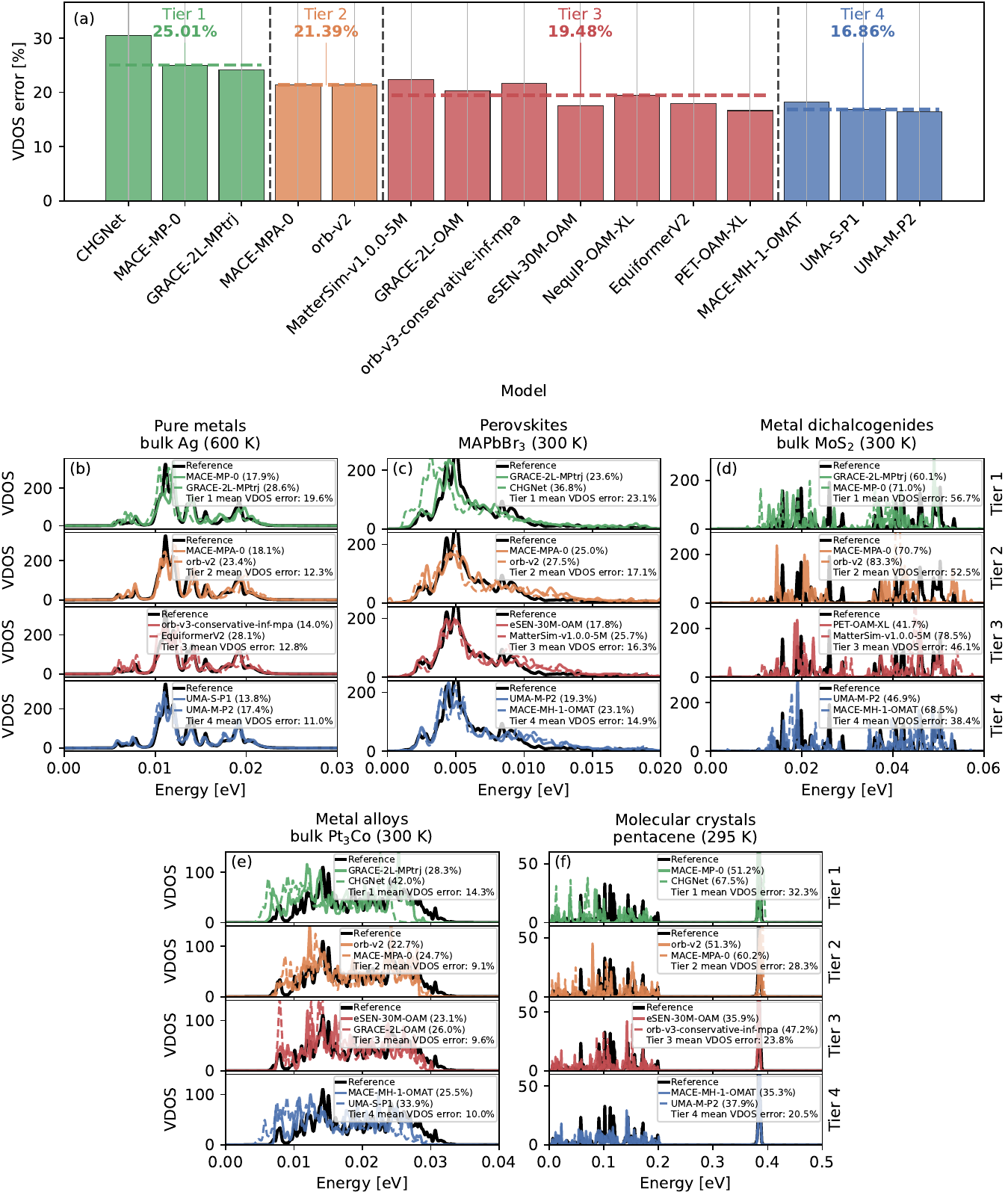}
    \caption{
    Vibrational density of states panel.
    (a): VDOS error for the MLIPs grouped in four tiers. The horizontal dashed lines indicate median values of the VDOS error for the four tiers.
    (b--f): reference and predicted normalised VDOS plots of the worst-performing systems for b) pure metals, c) perovskites, d) metal dichalcogenides, e) metal alloys, and f) molecular crystals.}
    \label{fig:plot-vdos-panel}
\end{figure*}

\begin{figure*}[t]
    \centering
    \includegraphics[width=1\linewidth]{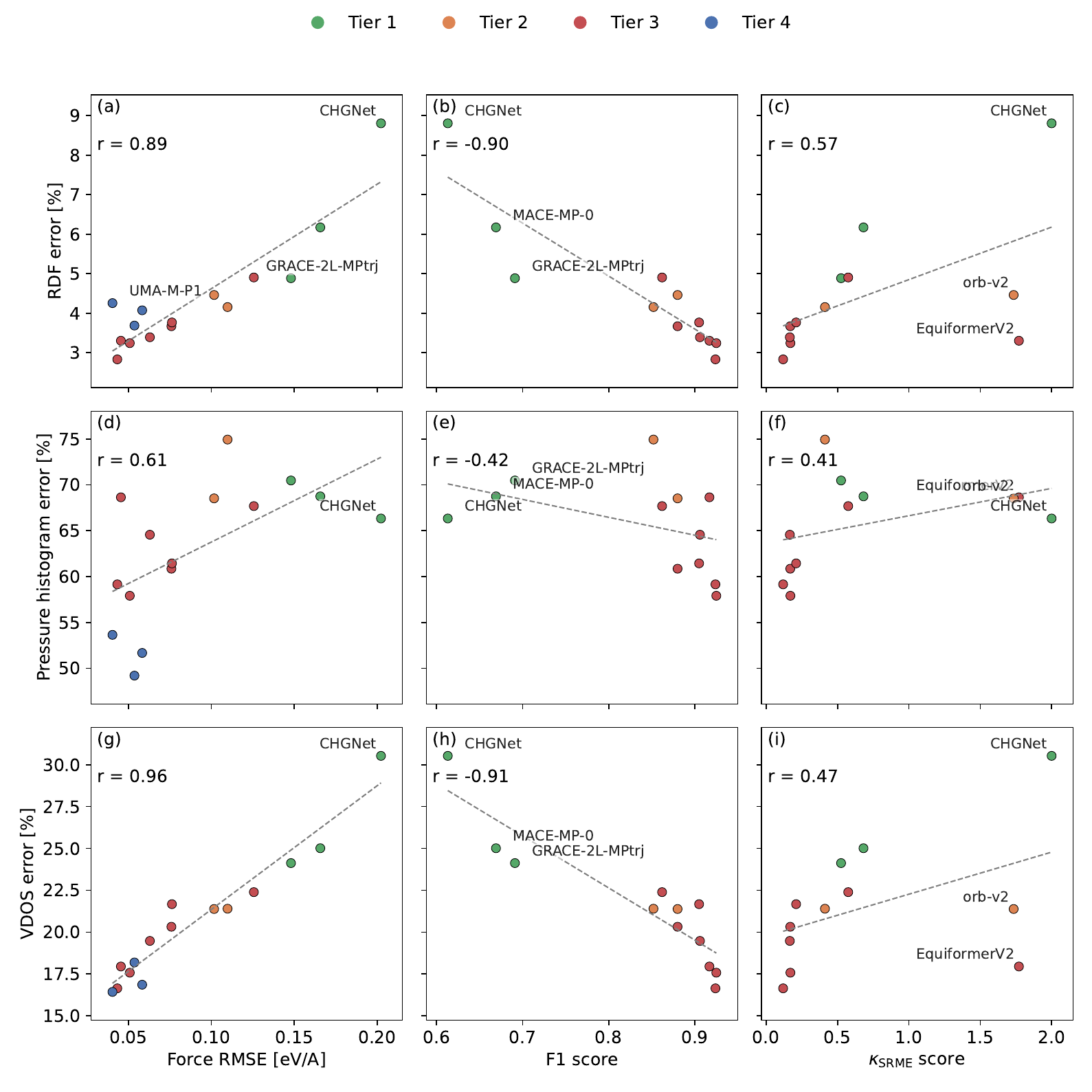}
    \caption{(a--c) RDF error, (d--f) pressure histogram error, and (g--i) VDOS error correlation plots. In each subplot, the value of the Pearson correlation $r$ is given in the top left corner, and the names of only the outliers are shown.
    Each observable error is correlated with respect to the force RMSE, F1 score, and $\kappa_{\text{SRME}}$, respectively, where the latter two are read off from the Matbench-discovery leaderboard. Tier-4 MLIPs are included only in the force RMSE correlation subplots. Refer to Figs~\ref{fig:plot-SI-rdf-correlations}--\ref{fig:plot-SI-vdos-correlations} for the correlation analysis with all model names shown.}
    \label{fig:plot-rdf-vdos-correlation}
\end{figure*}

\noindent To understand whether particular material classes dominate this behaviour, we next examine the pressure distributions and errors by material type.
As shown in Fig.~\ref{fig:plot-pressure-panel}(b-e), the largest pressure MAEs are obtained for metal alloys, with Tier 1 achieving the lowest mean pressure MAE of 2.75~GPa.
Tiers~2 and 3 perform similarly, attaining mean pressure MAEs of approximately 5~GPa, while Tier~4 exhibits an anomalously large mean pressure MAE of 28.20~GPa due to the poor performance of the \texttt{UMA-M-P1} model on bulk Pt$_3$Co.
We note that this system also yields an outlier RDF error for the same models.
Although the unit cell used for Pt$_3$Co corresponds to ambient conditions observed experimentally, it represents a highly out-of-equilibrium configuration for the \texttt{UMA-M-P1} model (the same is true for \texttt{UMA-S-P1}).
As shown in Fig.~\ref{fig:plot-SI-pressure-pt3co-uma-s-relaxed-cell} for \texttt{UMA-S-P1}, reasonable-pressure $NVT$ simulations for this system can be recovered by first relaxing the cell at zero pressure. 
This highlights the need for caution when using experimentally determined unit cells directly, without first relaxing the cell with the model being tested.
For all other material classes, the mean pressure MAE remains below 2~GPa across all tiers. In the case of pure metals, Tier 4 achieves the lowest mean pressure MAE of 0.22~GPa, with \texttt{UMA-M-P1} reaching a pressure MAE of only 0.1~GPa.
The lowest pressure MAEs are observed for perovskites and TMDs. For TMDs, Tier-4 models perform best, achieving a mean pressure MAE of 0.25~GPa, with \texttt{UMA-S-P1} reaching a sub-0.1~GPa pressure MAE.
For perovskites, all tiers achieve sub-1~GPa mean pressure MAEs, although there is no clear improvement with increasing model tier. In this case, Tier 1 obtains the lowest mean pressure MAE of 0.19~GPa.
We also quantify pressure prediction using the error metric defined in Eq.~(\ref{eq:pressure-histogram-error}), which quantifies the error based on the overlap of the reference and MLIP pressure histograms.
In Fig.~\ref{fig:plot-SI-pressure-histogram-error-panel}, we report the pressure errors for all MLIPs and highlight the systems with the largest mean errors.
We note that this metric does not discriminate well between very poor predictions. 
Pressure distributions with little or no overlap with the reference all return errors close to 100\%, making the metric less sensitive to the exceptionally large pressure errors observed for \texttt{UMA-M-P1}.
Based on this distribution-overlap metric, Tier~4 achieves the lowest median pressure error, indicating the best overall agreement between the MLIP and reference pressure distributions.\\

\noindent The relatively poor pressure errors observed for most models may partly reflect the sensitivity of stress predictions to the quality of the underlying DFT stress labels. 
Several foundation MLIPs were trained on datasets generated with DFT settings optimised primarily for large-scale energy and force data generation, rather than for highly converged stress tensors. 
For example, \textsf{MPTrj} uses a very coarse \vaspkw{EDIFF\_PER\_ATOM = 5e-05} and version~52 pseudopotentials, while \textsf{OMat24} largely follows the \textsf{MPTrj} settings but uses version~54 pseudopotentials. 
By contrast, our reference trajectories use a tighter absolute electronic convergence threshold, \vaspkw{EDIFF = 1e-6} and pseudopotentials consistent with \textsf{OMat24}. 
The comparatively lower pressure errors obtained by \texttt{MACE-MH-1-OMAT} and the \texttt{UMA} models (except for the model-specific failure for Pt$_3$Co also reflected in the RDF), may be due to a greater consistency between \textsf{OMat24} stress labels and our reference pressures.
However, overall, the large pressure errors may suggest that the DFT settings used in current large-scale training datasets are sufficient for learning useful energies and forces, but not necessarily for stress tensors.

\subsubsection{Vibrational density of states}

\noindent We now turn to probing the dynamical response of a system at finite temperature.
Within linear response theory~\cite{kubo_statistical-mechanical_1957}, this is captured by the velocity autocorrelation function yielding the vibrational density of states (VDOS)~\cite{green_markoff_1954}. 
Similar to the pressure analysis, we quantify the VDOS error using the metric defined in Eq.~(\ref{eq:vdos-error}), which provides a single-valued measure for comparing MLIP trajectories against the \textsf{Dyna-Mat-v1.0} reference. \\

\noindent The mean VDOS error, averaged over all trajectories, is summarised in Fig.~\ref{fig:plot-vdos-panel}(a) and classified by model tier.
Compared with RDFs, the median VDOS errors are generally higher, which may partly reflect the non-negligible statistical uncertainty associated with estimating both reference and predicted VDOS from trajectories of only 20~ps.
The median VDOS error decreases from ~25.01\% by about 3.62\% going from Tier ~1 to Tier~2.
The accuracy gains from Tiers~2-4 are monotonic but smaller. \\

\noindent To understand whether the higher errors are uniformly spread or system dependent, we next examine representative VDOS predictions by material class.
In Fig.~\ref{fig:plot-vdos-panel}(b--f), we compare the best- and worst-performing models against the reference VDOS for five worst-performing systems, each belonging to a different material class.
The largest VDOS errors are observed for TMDs, where even the best-performing model, \texttt{PET-OAM-XL}, yields an error of 41.7\%, while the worst-performing model, \texttt{orb-v2}, reaches 83.3\%.
Molecular crystals also exhibit relatively large VDOS errors, with the mean VDOS errors ranging from $\sim$20\% up to $\sim$30\%. 
Possible explanations for the large errors in these systems include the presence of many sharply peaked bands over a wide frequency range, all of which must be accurately reproduced to obtain a low error, as well as limited statistical convergence of low-frequency collective modes. 
By contrast, pure metals, metal alloys, and to a lesser extent perovskites are described substantially better, with all model tiers achieving sub-25\% mean VDOS error and metal alloys showing the lowest mean errors overall.
Overall, these results show that errors in VDOS predictions are strongly system dependent.

\subsection{Assessment of metrics}

\noindent We now contextualise our error metrics with respect to those used in the current community benchmark leaderboard, Matbench-discovery.
Among the various metrics reported, the F1 score for classifying whether structures relaxed with an MLIP lie above or below the convex hull, and $\kappa_{\text{SRME}}$, which probes lattice thermal conductivity from phonons and third-order force constants, are increasingly used by the community to assess MLIP performance.
It is worth noting that the ground-truth DFT settings used for the F1 and $\kappa_{\mathrm{SRME}}$ benchmarks are consistent with \textsf{MPTrj} and \textsf{Alexandria}, but not with \textsf{OMat24}. 
As a result, Tier~4 models are not included in these leaderboards and are therefore excluded from this correlation analysis. \\

\noindent Often, the F1 score is regarded as a probe of energy accuracy, since forces enter only indirectly through structural relaxation, whereas $\kappa_{\text{SRME}}$ is considered a probe of force accuracy and is further linked to anharmonicity and finite-temperature behaviour.
Within this premise, one could think that $\kappa_{\text{SRME}}$ would correlate more strongly with the errors in VDOS and the RDF than the F1 score. 
However, we observe the opposite trend. 
As shown in Fig.~\ref{fig:plot-rdf-vdos-correlation}, the F1 score shows a much stronger (anti)correlation with the RDF and VDOS errors, averaged over all systems, whereas $\kappa_{\text{SRME}}$ shows only a moderate-to-weak correlation.
We also find that force RMSEs estimated on the \textsf{Dyna-Mat-v1.0} trajectories correlate strongly with the F1 score as well as with the RDF and VDOS errors, suggesting that it can be a useful metric for assessing MLIP performance at finite temperatures. \\

\noindent The strong correlation between observable errors and the F1 score indicates that, across the models considered here, improvements in the relative energies of local minima are accompanied by improvements in the shape and curvature of the potential energy surface.
However, because this correlation is evaluated across heterogeneous model generations, it may also partly reflect a common improvement trend in newer models, rather than a direct mechanistic relationship between F1 performance and finite-temperature observables.
Together, these results are consistent with a broad improvement in foundation MLIPs across the potential energy surface, rather than an improvement restricted to isolated local regions.
The weak correlation with $\kappa_{\mathrm{SRME}}$ is likely due to the noisy nature of this estimator, as noted from its sensitivity to the finite displacement used in phonon calculations~\cite{orb-v3}. \\

\noindent We further note that pressure errors correlate poorly with all metrics considered here, including the F1 score, $\kappa_{\mathrm{SRME}}$, force RMSEs, and even RDF and VDOS errors, as shown in Figs~\ref{fig:plot-SI-pressure-histogram-correlations} and \ref{fig:plot-SI-correlation-rdf-pressure-same-length}--\ref{fig:plot-SI-correlation-pressure-vdos-same-length}.
This decoupling indicates that pressure prediction is limited by a distinct source of error from that governing structural and dynamical observables. 
A possible explanation is noisy or insufficiently converged stress labels in large-scale training datasets as a likely bottleneck.
Given this decoupling, structural and dynamical error metrics are insufficient proxies for pressure accuracy. 
Therefore, pressure must be validated independently.
Finally, in Fig.~\ref{fig:plot-SI-correlation-rdf-vdos-same-length} we show the strong correlation of the RDF and VDOS errors.

\begin{figure*}[t]
    \centering
    \includegraphics[width=\linewidth]{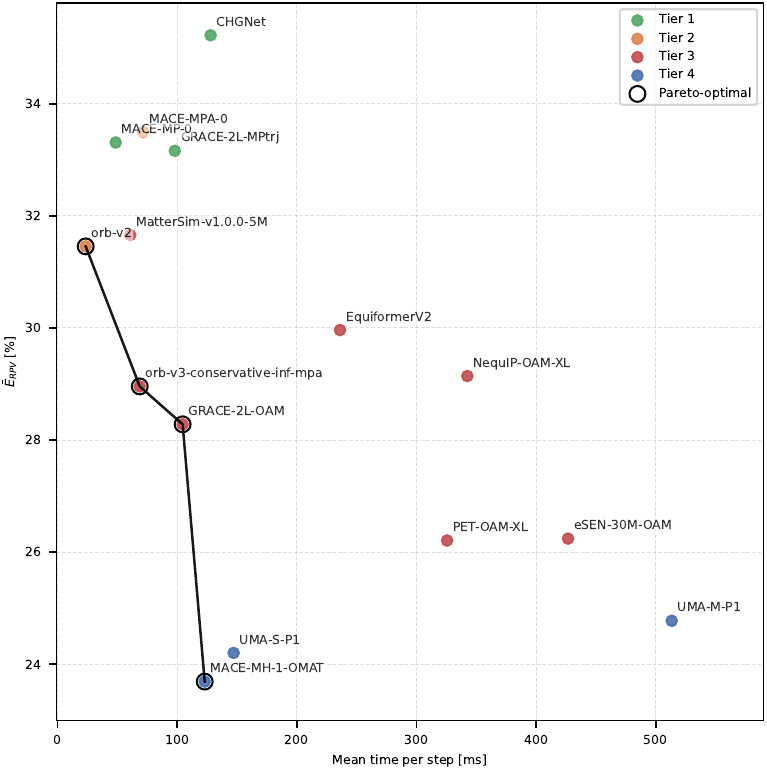}

    \caption{Pareto front of the combined error metric and the mean time per MD step using ASE Calculators (single precision for the MLIP forwards without acceleration libraries). The colours denote four model tiers, and the Pareto-optimal MLIPs are indicated with the black circle.}
    \label{fig:plot-pareto-vdos-rdf}
\end{figure*}

\subsection{Pareto front}

\noindent Finally, having assessed a range of finite-temperature properties using models of different sizes and computational costs, we summarise the overall performance of foundation MLIPs by constructing an accuracy--cost Pareto frontier.
To facilitate this analysis, we define an aggregate accuracy metric that combines the three observable-level error metrics introduced above.
Fig.~\ref{fig:plot-pareto-vdos-rdf} shows the position of all MLIPs considered in this work relative to the Pareto frontier, using the aggregate accuracy metric and the mean time per MD step measured on an NVIDIA V100 GPU with 16 GB RAM on the GPU, through the ASE calculator interface.
We note that it is easy to conflate the implementation of MLIPs  with the intrinsic cost of the architectures, as implemented in development codebases such as PyTorch. 
This can arise because acceleration libraries may speed up equivariant tensor products, while certain codes only allow single-precision calculations. 
Thus, to impose as much fairness as possible when assessing relative costs, we disable all acceleration libraries and use float32. \\

\noindent The Pareto frontier consists of three models.
The first Pareto-optimal model is the Tier~2 \texttt{orb-v2}, followed by the Tier~3 \texttt{GRACE-2L-OAM}.
The final Pareto-optimal model is the Tier~4 \texttt{MACE-MH-1-OMAT}, which achieves the lowest $\bar{E}_{\mathrm{RPV}}$ error (see Eq.~(\ref{eq:combined-metric})) overall while retaining a sub-150~ms mean time per MD step.
To assess the sensitivity of the Pareto-optimal set to the definition of the aggregate accuracy metric, we also report in Fig.~\ref{fig:plot-pareto-vdos-rdf-panel} the Pareto frontier obtained using the alternative metric defined in Eq.~(\ref{eq:combined-metric-2}), in which the individual errors are combined with weights that scale linearly with their magnitude, thereby assigning greater weight to the observables for which a model performs poorly.
The corresponding Pareto frontiers for RDF, pressure histogram, and VDOS errors are reported separately in Figs~\ref{fig:plot-SI-pareto-rdf-time}--\ref{fig:plot-SI-pareto-vdos-time}.

\section{Conclusions}

\noindent In summary, we introduce \textsf{Dyna-Mat-v1.0}, a benchmark dataset of condensed-phase first-principles MD trajectories designed to assess the performance of foundation MLIPs under finite-temperature conditions.
Using \textsf{Dyna-Mat-v1.0}, we benchmark 15 foundation MLIPs across four model tiers.
We first evaluate single-point energy and force RMSEs on configurations sampled from the reference trajectories, and then assess finite-temperature performance by running MLIP-driven molecular dynamics with matched simulation settings and comparing RDFs, pressure distributions, and VDOS against the corresponding first-principles results. 
We define normalised error metrics for these observables, scaled from 0 to 100\%, and combine them into a single aggregate error score.
Finally, we construct an accuracy--cost Pareto frontier to assess the trade-off between finite-temperature accuracy and computational efficiency. \\

\noindent Across the models considered here, single-point errors evaluated on the \textsf{Dyna-Mat-v1.0} trajectories decrease systematically from Tier~1 to Tier~4. 
This trend suggests that both the increased training-set size from Tier~1 to Tier~2 and the inclusion of \textsf{OMat24}-derived data in Tier~3 reduce errors on finite-temperature configurations. 
Tier~4 models yield the lowest single-point errors overall, which may reflect either closer consistency between the DFT settings used in \textsf{OMat24}-based training data and those used in \textsf{Dyna-Mat-v1.0}, relative to \textsf{MPTrj} and \textsf{Alexandria}, or the substantially larger scale of \textsf{OMat24} compared with these earlier datasets.
The errors on observables, however, show a more interesting picture. 
The tier-median errors on RDFs are consistently smaller than 10 \%, and exhibit diminishing returns Tier 2 onward, with Tier 3 models surprisingly yielding lower errors. 
This is due to UMA models exhibiting a failure for Pt$_3$Co at 300\,K. 
The pressure distributions reveal a potential issue with DFT data quality, with most models trained or fine-tuned on \textsf{MPTrj} or \textsf{Alexandria} yielding non-negligible MAEs.
Models trained on \textsf{OMat24} generally yield lower pressure errors, apart from the outlier behaviour of the \texttt{UMA} models for Pt$_3$Co at 300\,K, where the error exceeds 1000\%.
This suggests future material datasets should consider stricter DFT convergence settings. 
And finally, we consider a dynamical property the vibrational density of states and in this case note a systematic improvement in the error metric from approximately 25 \% to approximately 17 \% across model Tiers. \\

\noindent We also compare the finite-temperature errors obtained from \textsf{Dyna-Mat-v1.0} with metrics used in existing community benchmarks.
Force RMSEs evaluated directly on the \textsf{Dyna-Mat-v1.0} trajectories correlate strongly with RDF and VDOS errors, indicating that force errors on finite-temperature configurations provide a useful diagnostic of MD reliability for structural and dynamical properties.
These trends should not, however, be interpreted solely in terms of model tier: the tier axis also partly conflates training-set size with training-data composition.
In particular, whether a model has been trained on MD or finite-temperature configurations, such as first-principles MD-like data in \textsf{OMat24}, rather than primarily on relaxed structures, may be an important factor in determining its finite-temperature accuracy.
By contrast, the correlation breaks down for pressure distributions, reinforcing that thermodynamic stress is a distinct and more demanding test of MLIP accuracy.
In contrast, among the external benchmark metrics considered, the Matbench-discovery F1 score correlates more strongly with RDF and VDOS errors than $\kappa_{\mathrm{SRME}}$, despite the latter being more directly associated with phonons, anharmonicity, and thermal transport.
This suggests that no single existing benchmark metric fully captures finite-temperature transferability, and that direct validation on thermodynamic ensembles remains necessary. \\

\noindent When finite-temperature accuracy is considered together with inference cost, the picture is more nuanced.
Although the median accuracy gains beyond Tier~2 are often modest, several cross-trained Tier~4 models lie on, or close to, the Pareto frontier.
Thus, the latest multi-dataset models can offer favourable accuracy--efficiency trade-offs for MD applications, even when their average errors do not uniformly outperform all lower-tier models across every observable. \\

\noindent We conclude by noting several limitations and areas for future work.
Although \textsf{Dyna-Mat-v1.0} contains a large number of configurations, it is built from 17 first-principles trajectories and includes only a small number of systems within each material class.
In addition, each system is represented by a single reference trajectory, so the benchmark does not explicitly sample trajectory-to-trajectory variability arising from different initial conditions.
The simulations are also restricted to fixed-cell dynamics, and therefore do not assess MLIP performance for variable-cell finite-temperature simulations or cell-fluctuation-driven properties.
Furthermore, the present benchmark focuses on fundamental structural, thermodynamic, and dynamical observables.
These aspects can be extended naturally by adding further chemistries, phases, thermodynamic conditions, multiple independent trajectories per system, variable-cell simulations, and more complex finite-temperature properties.
Taken together, these results establish \textsf{Dyna-Mat-v1.0} as a reusable reference benchmark for assessing foundation MLIPs beyond static and harmonic properties, and provide practical guidance for selecting and validating MLIPs for finite-temperature atomistic simulations.

\section{Methods}
\label{sec:methods}

\subsection{Molecular dynamics}

\subsubsection{First-principles molecular dynamics}

\begin{figure*}[t]
    \centering
    \includegraphics[width=\linewidth]{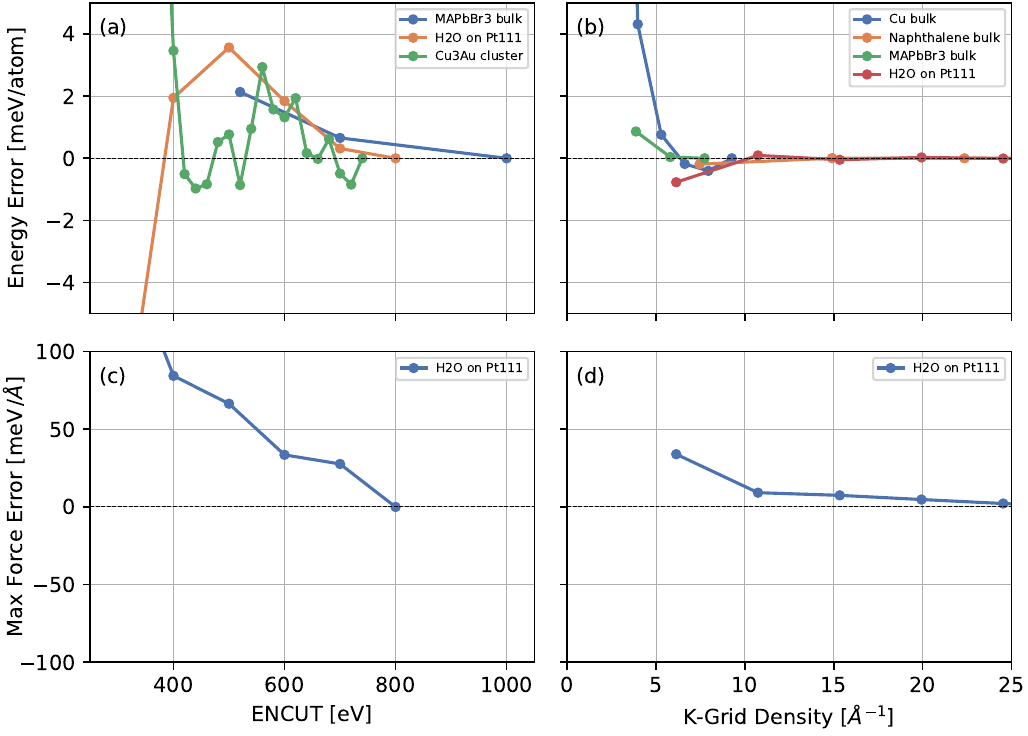}
    
    \caption{Convergence tests on our dataset; a) Convergence of the total energy against the plane-wave cutoff (ENCUT); b) Convergence of the total energy against the k-grid density; c) Convergence of the maximal force against the plane-wave cutoff (ENCUT); d) Convergence of the maximal force against the k-grid density.}
    \label{fig:convergence_tests}
\end{figure*}

\noindent We performed first-principles molecular-dynamics using the PBE exchange correlation functional~\cite{perdew_generalized_1996}. D3~\cite{GrimmeBannwarth2016} dispersion interactions were added for the molecular crystal simulations.
The DFT computational settings used in production simulations are summarised in Table~\ref{tab:dft_settings_comparison}.
To ensure numerical accuracy and cross-platform consistency, systematic convergence benchmarks were performed to establish the plane-wave cutoff, $k$-grid density, and parameters between VASP and FHI-aims (Fig.~\ref{fig:convergence_tests}). 
As shown in Fig.~\ref{fig:convergence_tests}(a), the per-atom-energy error for bulk systems was evaluated against a 1000~eV reference. A lower cutoff, of below 400~eV, exhibits a large truncation error ($>$5.0~meV/atom), whereas 700~eV significantly reduces this error to $<$2.0~meV/atom. Based on the convergence profile of both energies and forces, an \texttt{ENCUT} of 700~eV was selected to balance accuracy with computational efficiency.
The corresponding force convergence for the $\text{H}_2\text{O}$/Pt(111) interface is shown in Fig.~\ref{fig:convergence_tests}(c). At low cutoff energies, the maximum force error remains significant ($\sim$80.0~meV\AA$^{-1}$ at 400~eV), indicating insufficient basis-set completeness for reliable structural optimisation. Increasing the cutoff systematically suppresses the force error, reaching a force error of $<$50~meV\AA$^{-1}$ near 700~eV. This confirms that the selected \texttt{ENCUT} provides sufficiently converged atomic forces for molecular dynamics simulations.
Reciprocal space sampling was benchmarked against the $k$-grid density ($\text{\AA}^{-1}$) for four distinct material classes (Fig.~\ref{fig:convergence_tests}(b)). 
The plane-wave cutoff was chosen to be 700~eV.
Above our minimum k-grid density of 4.54~$\text{\AA}^{-1}$, corresponding to \texttt{KSPACING = 0.22} in VASP, energy errors fall roughly below 2.0~meV/atom. A k-grid density of $\geq$4.54~$\text{\AA}^{-1}$ was selected for all production calculations.
Fig.~\ref{fig:convergence_tests}(d) demonstrates the force convergence with respect to $k$-grid density for the $\text{H}_2\text{O}$/Pt(111) system. The force error remains low across the investigated sampling range, indicating that the atomic forces are substantially less sensitive to reciprocal-space discretisation than the total energy. Nevertheless, the chosen $k$-grid density of 4.54~$\text{\AA}^{-1}$ ensures consistent convergence of both electronic energies and ionic forces, particularly for metallic systems.
Final settings for VASP and FHI-aims are shown in Table~\ref{tab:dft_settings_comparison}. By strictly aligning these parameters, we ensure that both the plane-wave pseudopotential and all-electron frameworks are rigorously converged to a common baseline. Such well-converged settings have been shown across community-wide benchmarks to eliminate code-specific numerical variations, yielding highly reproducible and virtually identical results for solid-state properties \cite{lejaeghere2016reproducibility}.
\begin{table*}[htbp]
  \centering
  \caption{Comparison of DFT settings between VASP and FHI-aims}
  \label{tab:dft_settings_comparison}

  \begin{tabular}{l|c|c}
    \hline
    \textbf{Setting} & \textbf{VASP} & \textbf{FHI-aims} \\
    \hline

    Basis set / cutoff 
    & 700 eV 
    & \texttt{tight} (2020 defaults) \\
    \hline

    Core treatment 
    & \texttt{potpaw.64} 
    & all-electrons \\
    \hline

    k-point sampling 
    & $4.54~\text{\AA}^{-1}$ ($\Gamma$-centered)
    & $4.54~\text{\AA}^{-1}$ ($\Gamma$-centered)
    \\
    \hline

    Electronic smearing 
    & Gaussian, $\sigma = 0.05$ eV 
    & Gaussian, $\sigma = 0.05$ \\
    \hline

    Dipole correction (surfaces) 
    & Enabled 
    & Enabled \\
    \hline

    Cluster setup 
    & Periodic supercell with vacuum 
    & Non-periodic isolated cluster \\
    \hline

    Symmetry 
    & Symmetry disabled 
    & Symmetry disabled \\
    \hline

  \end{tabular}
\end{table*}

\subsubsection{\texorpdfstring{$\mathrm{NVT}$}{NVT} with MLIPs}
\noindent The simulations for pure metals, perovskites, TMDs, and metal alloys are performed using ASE~\cite{ASE} in the $\mathrm{NVT}$ ensemble with MLIP interatomic potentials. These simulations are thermalised using a single Nos\'e--Hoover chain thermostat, a time step of 0.25 fs, and a thermostat characteristic time scale of 254.5 fs. Positions and potential energies are sampled every 10 time steps over a total simulation time of 20 ps.
The MD simulations for molecular crystals are performed using \texttt{i-PI}~\cite{litman_i-pi_2024} in the $\mathrm{NVT}$ ensemble with MLIPs. These simulations are thermalised using a Langevin thermostat with a time step of 0.5 fs. The thermostat time constant is set to 10 fs for anthracene, naphthalene, and tetracene, and to 500 fs for pentacene and picene. For all molecular crystals, positions and potential energies are sampled every time step over a total simulation time of 20 ps.

\subsection{Error metrics}

\subsubsection{Single-point energy and forces}
\noindent To test the MLIPs on \textsf{Dyna-Mat-v1.0} trajectories, we estimate single-point energy and force root mean square error (RMSE) as defined below
\begin{align}
    U_{\text{RMSE}} &= \sqrt{\frac{1}{MN^2}\sum_{i=1}^M{\left(U^{\text{MLIP}}_i - U^{\text{REF}}_i\right)^2}},\label{eq:energy-rmse}\\
    F_{\text{RMSE}} &= \sqrt{\frac{1}{MN}\sum_{i=1}^M{\left\|\mathbf{F}^{\text{MLIP}}_{i} - \mathbf{F}^{\text{REF}}_{i}\right\|_2^2}},\label{eq:force-rmse}
\end{align}
where $M$ is the number of configurations, $N$ is the number of atoms, $\{U^{\text{MLIP}}_i, U^{\text{REF}}_i\}^{M}_{i=1}$ are the predicted and reference total energies, and $\{\mathbf{F}^{\text{MLIP}}_{i}, \mathbf{F}^{\text{REF}}_{i}\}^{M}_{i=1}$ are the predicted and reference force vectors for all atoms in configuration $i$.

\subsubsection{Radial distribution functions}

\noindent To probe the structural properties of the systems under consideration, we first compute the radial distribution function,
\begin{equation}
    g(r)=\frac{1}{4\pi r^2\rho N}
    \left < \sum_{j=1}^N\sum_{\substack{j'=1\\j'\neq j}}^N \delta(r-r_{jj'}) \right >,
    \label{eq:rdf-def}
\end{equation}
where $r$ is the distance from a reference particle, $\rho$ is the particle density, $N$ is the number of particles, and $\left < \cdot \right >$ denotes an ensemble average.
For each system, we compute the total RDF over all unique pairs of atoms for both the reference and MLIP trajectories. Pair distances are accumulated into 500 radial bins, and $R_{\text{max}}$ is taken as half the shortest cell vector. All reference and MLIP RDFs are computed using the MDTraj package~\cite{MDTraj}.
We then quantify the difference between the reference and MLIP RDFs using the RDF error $E_{\text{RDF}}$, defined as
\begin{equation}
\begin{aligned}
E_{\mathrm{RDF}} &= \min\\
&\left(1,
\frac{\int_0^{R_{\max}} \left| g^{\mathrm{REF}}(r) - g^{\mathrm{MLIP}}(r) \right| \, dr}
     {\int_0^{R_{\max}} \left| g^{\mathrm{REF}}(r) - 1 \right| \, dr}\right)\\
     &\times100\%,
\end{aligned}
\label{eq:rdf-error}
\end{equation}
where $g^{\mathrm{REF}}(r)$ and $g^{\mathrm{MLIP}}(r)$ are the RDFs obtained from the reference and MLIP trajectories, respectively.
For each model, RDFs are computed for all systems over matched-duration trajectory segments and compared with the corresponding reference RDFs to obtain the mean RDF error. The metric $E_{\text{RDF}}$ measures the deviation of the MLIP RDF from the reference RDF; a perfect match gives $E_{\text{RDF}}=0\%$.
The $\left|g^{\mathrm{REF}}(r)-1\right|$ term normalises the error by the deviation of the reference RDF from the ideal-gas RDF. If this deviation is small, the normalised error becomes sensitive to small differences and may saturate at 100\%. If the denominator is exactly zero, we assign an RDF error of 100\%.

\subsubsection{Pressure}

\noindent To probe the equilibrium densities, we report the error on the pressure. 
Given the virial tensor $\boldsymbol{\sigma}$, we compute the pressure as
\begin{equation}
P =
\begin{cases}
-\dfrac{1}{3}
\left(\sigma_{xx}+\sigma_{yy}+\sigma_{zz}\right),
& \text{for 3D systems}, \\[6pt]
-\dfrac{1}{2}
\left(\sigma_{xx}+\sigma_{yy}\right),
& \text{for 2D systems}.
\end{cases}
\end{equation}
For each system $s$ and MLIP model $m$, the reference and MLIP
trajectories are truncated to matched simulation durations. Since the
trajectories may use different time steps, their pressures are averaged
separately:
\begin{equation}
\begin{aligned}
\overline{P}^{\mathrm{REF}}_s &=
\frac{1}{M^{\mathrm{REF}}_s}
\sum_{i=1}^{M^{\mathrm{REF}}_s} p^{\mathrm{REF}}_{s,i}, \\
\overline{P}^{\mathrm{MLIP}}_{m,s} &=
\frac{1}{M^{\mathrm{MLIP}}_{m,s}}
\sum_{i=1}^{M^{\mathrm{MLIP}}_{m,s}} p^{\mathrm{MLIP}}_{m,s,i}.
\end{aligned}
\end{equation}
The pressure MAE for a given system is
\begin{equation}
\Delta P_{m,s} =
\left|
\overline{P}^{\mathrm{REF}}_s -
\overline{P}^{\mathrm{MLIP}}_{m,s}
\right|.
\label{eq:pressure-mae-error-per-system}
\end{equation}
To obtain a condensed performance score for each MLIP, we report the
mean pressure error across the $S$ systems:
\begin{equation}
\Delta P_m =
\frac{1}{S}
\sum_{s=1}^{S} \Delta P_{m,s}.
\label{eq:pressure-mae-error-per-model}
\end{equation} \\
We also consider a pressure-distribution error based on the overlap of the
pressure histograms. For each system and MLIP, we construct area-normalised
reference and MLIP pressure histograms using shared bin edges. Given the
histogram densities
$\{h_k^{\mathrm{REF}},h_k^{\mathrm{MLIP}}\}_{k=1}^{K}$, we define the pressure histogram error:
\begin{equation}
E_P =
\frac{
\sum_{k=1}^{K}
\left|h_k^{\mathrm{REF}}-h_k^{\mathrm{MLIP}}\right|\Delta_k
}{
\sum_{k=1}^{K} h_k^{\mathrm{REF}}\Delta_k
+
\sum_{k=1}^{K} h_k^{\mathrm{MLIP}}\Delta_k
}
\times 100\%,
\label{eq:pressure-histogram-error}
\end{equation}
where $\Delta_k$ is the width of histogram bin $k$. For each MLIP, we report
the mean value of $E_P$ across systems. 
Note that the denominator of Eq.~(\ref{eq:pressure-histogram-error}) is identically equal to 2 before multiplication by 100\%. 

\subsubsection{Vibrational density of states}

\noindent  Finally, to investigate the dynamical properties, we compute the vibrational
density of states from the Hann-windowed velocity autocorrelation
functions. Before comparison, the reference and MLIP VDOS are matched in terms of the simulation lengths and normalised to unit area. To quantify their differences, we use the VDOS error, given by
\begin{equation}
\begin{aligned}
    E_{\mathrm{VDOS}} &=
    \frac{
        \int_{\Omega}
        \left| \tilde{f}^{\mathrm{REF}}(\nu)
        - \tilde{f}^{\mathrm{MLIP}}(\nu) \right| d\nu
    }{
        \int_{\Omega} \tilde{f}^{\mathrm{REF}}(\nu) d\nu
        + \int_{\Omega} \tilde{f}^{\mathrm{MLIP}}(\nu) d\nu
    } \\
    &\times 100\%,
    \label{eq:vdos-error}
    \end{aligned}
\end{equation}
where $\tilde{f}^{\mathrm{REF}}(\nu)$ and
$\tilde{f}^{\mathrm{MLIP}}(\nu)$ are the normalised reference and MLIP VDOS,
respectively, and $\Omega$ is their common sampled frequency range. For each
model, we report the mean $E_{\mathrm{VDOS}}$ averaged over all systems.

\subsubsection{Combined}
\noindent To facilitate the assessment of the model performance on the three metrics, namely RDF, pressure, and VDOS, we introduce the combined error score given by
\begin{equation}
    \bar{E}_{\text{RPV}} = \frac{E_{\text{RDF}}+E_{P}+E_{\text{VDOS}}}{3},
    \label{eq:combined-metric}
\end{equation}
which is an average of the three errors.\\
To penalise the poor performance on the pressure prediction, we introduce another combined error metric, based on the Lehmer mean of the errors
\begin{equation}
    L_{\text{RPV}} =\frac{E_{\text{RDF}}^2+E_{P}^2+E_{\text{VDOS}}^2}{E_{\text{RDF}}+E_{P}+E_{\text{VDOS}}}.
    \label{eq:combined-metric-2}
\end{equation}

\section{Acknowledgements}
This article is a result of joint work in COST Action CA22154 - Data-driven Applications towards the Engineering of functional Materials: an Open Network (DAEMON) supported by COST (European Cooperation in Science and Technology).
We thank the participants of the CECAM Flagship Workshop at Cornell Tech, Roosevelt Island, New York City on ``Physics-aware Machine Learning for Molecules and Materials'' for providing valuable feedback on these results presented at the meeting. We are especially grateful to Mit Kotak for highlighting the limitations of the performance-efficiency benchmarks used in an earlier version of the paper and for suggesting alternative approaches.
We thank Brandon Wood for bringing to our attention, and sharing molecular dynamics data illustrating, the improved pressure-histogram performance of the \texttt{UMA} model upon cell relaxation for Pt$_3$Co.  
JS was funded by the SNSF Ambizione grant number 233444.
SB acknowledges support from the National Science Centre, Poland, through the SONATA BIS grant DEC-2023/50/E/ST3/00569 and from the Foundation for Polish Science through the project FENG.02.02-IP.05-0177/23, co-financed by the European Union under the European Funds for Smart Economy 2021–2027. JK acknowledges that AWS resources were provided by the National Infrastructures for Research and Technology (GRNET) and funded by the EU Recovery and Resilience Facility under the DataMind project (555141862428).
The Flatiron Institute is a division of the Simons Foundation. LH was supported by a UKRI Horizon grant (MSCA, EP/Y024923/1). 

\section{Author Contributions}
Conceptualization: LH, VK;
Methodology: MJG, JS, IL, JR, LH, VK; 
Data generation: NA, HHH, JK, IL, MR, JS, SS, AW, BXS; 
Analysis: MJG, NA, SB, ASG, HHH, JK, IL, HM, JR, MR, MRu, JS, SS, BXS, AW, LH, VK; 
Data Curation: MJG, LH; 
Writing – Original Draft: MJG, VK; 
Writing – Review \& Editing: MJG, ASG, LH, MR, IL, HHH, JR, VK; 
Visualization: MJG, LH; 
Supervision: LH, VK; 
Project Administration: LH, VK; 

\section{Competing Interests}
The authors have no competing interests.

\section{Data and Code Availability}

\noindent The data and code will be made available in a public repository upon publication. In the meantime, they can be accessed through \href{https://matbench-discovery.materialsproject.org/}{Matbench-discovery}.
\clearpage
\onecolumngrid

\section*{Supporting Information}

\setcounter{figure}{0}
\renewcommand{\thefigure}{S\arabic{figure}}
\renewcommand{\theHfigure}{S\arabic{figure}}
\setcounter{table}{0}
\renewcommand{\thetable}{S\arabic{table}}
\renewcommand{\theHtable}{S\arabic{table}}

\section*{Single-point energy and force RMSEs}

\begin{figure}[H]
    \centering
    \includegraphics[width=\linewidth]{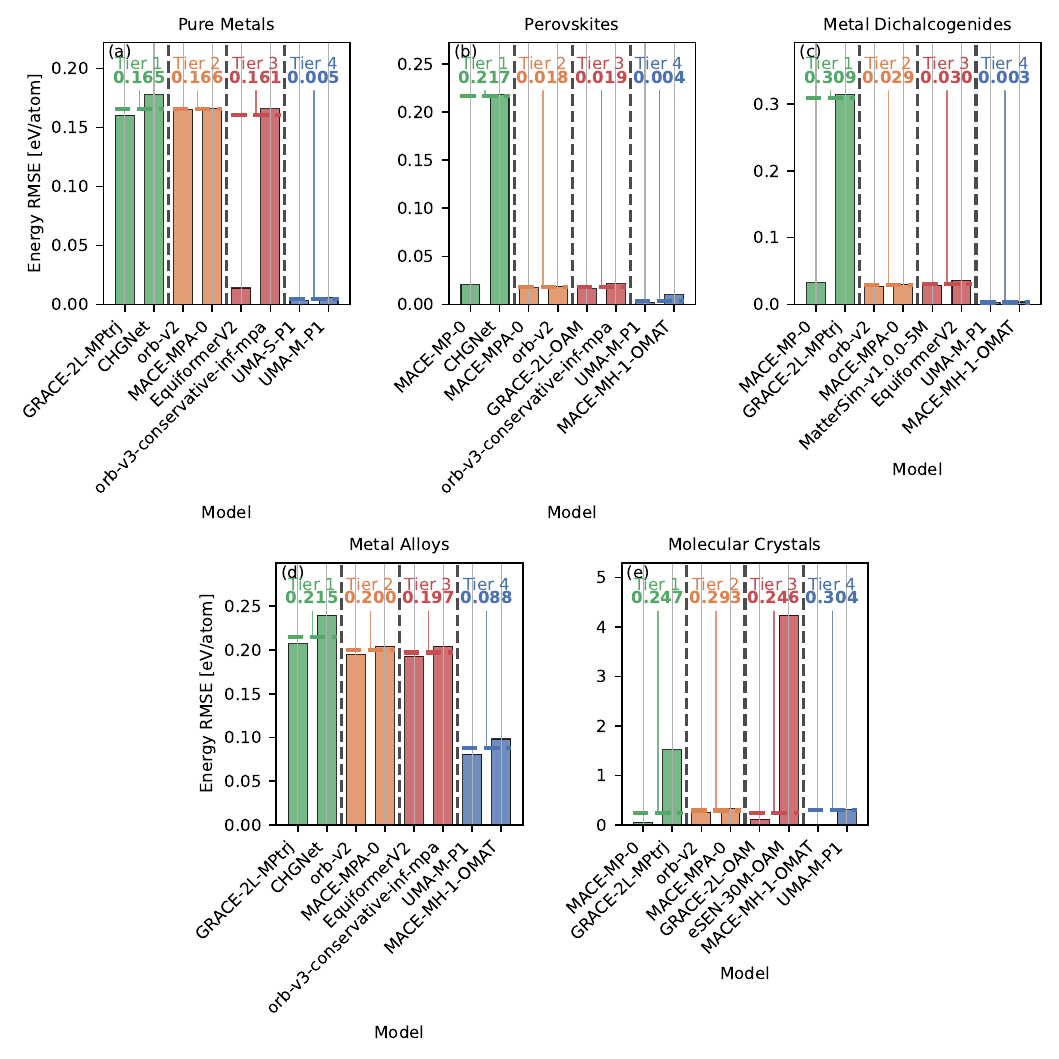}
    \caption{Energy RMSEs for the best and worst models from each model tier given for five different system types: a) pure metals, b) perovskites, c) metal dichalcogenides, d) metal alloys, and e) molecular crystals (does not include \texttt{EquiformerV2}). In each subplot, the tier-wise median energy RMSE is indicated with a horizontal dashed line.}
    \label{fig:plot-SI-energy-rmse}
\end{figure}

\newpage
\begin{figure}[H]
    \centering
    \includegraphics[width=\linewidth]{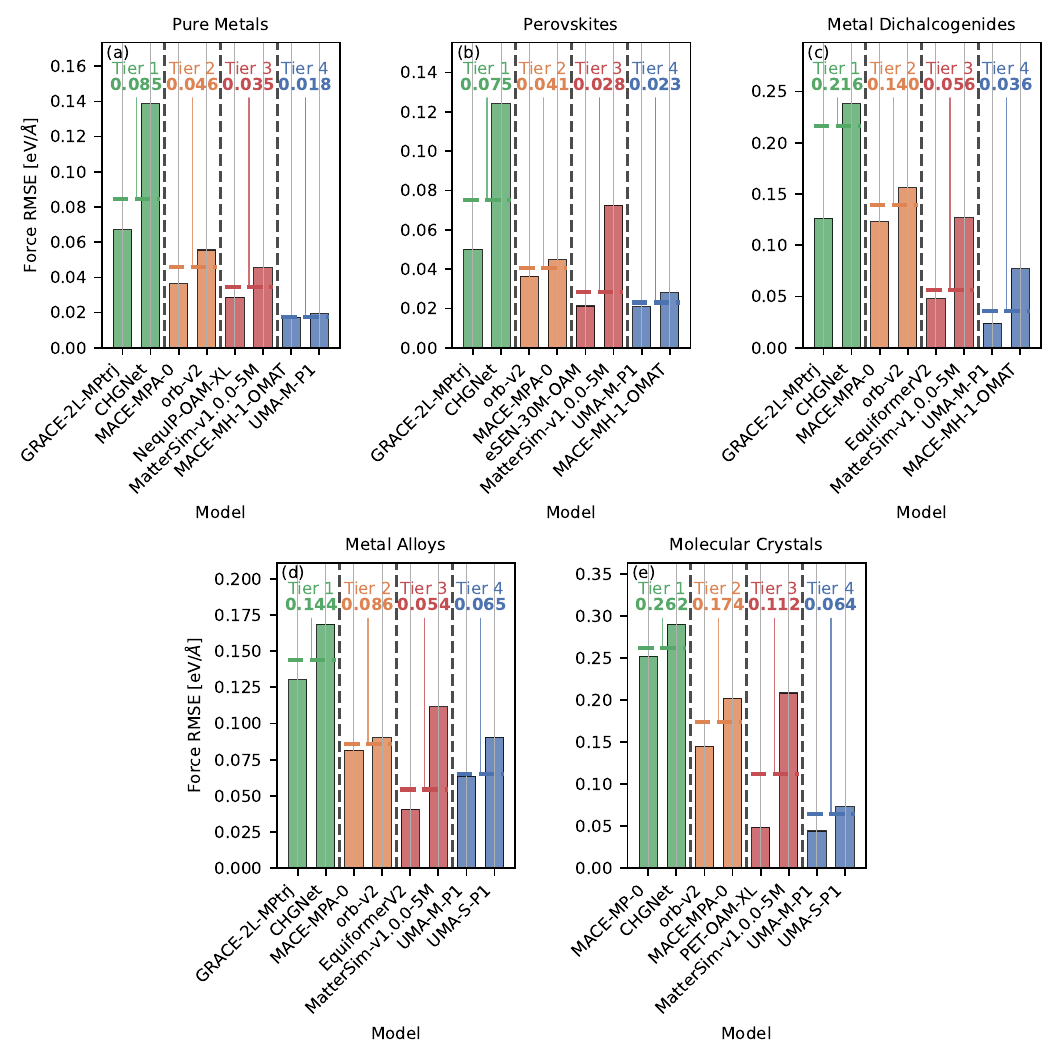}
    \caption{Force RMSEs for the best and worst MLIPs from each model tier given for five different system types: a) pure metals, b) perovskites, c) metal dichalcogenides, d) metal alloys, and e) molecular crystals. In each subplot, the tier-wise median force RMSE is indicated by a horizontal dashed line.}
    \label{fig:plot-SI-force-rmse}
\end{figure}

\newpage

\begin{figure}[H]
    \centering
    \includegraphics[width=\linewidth]{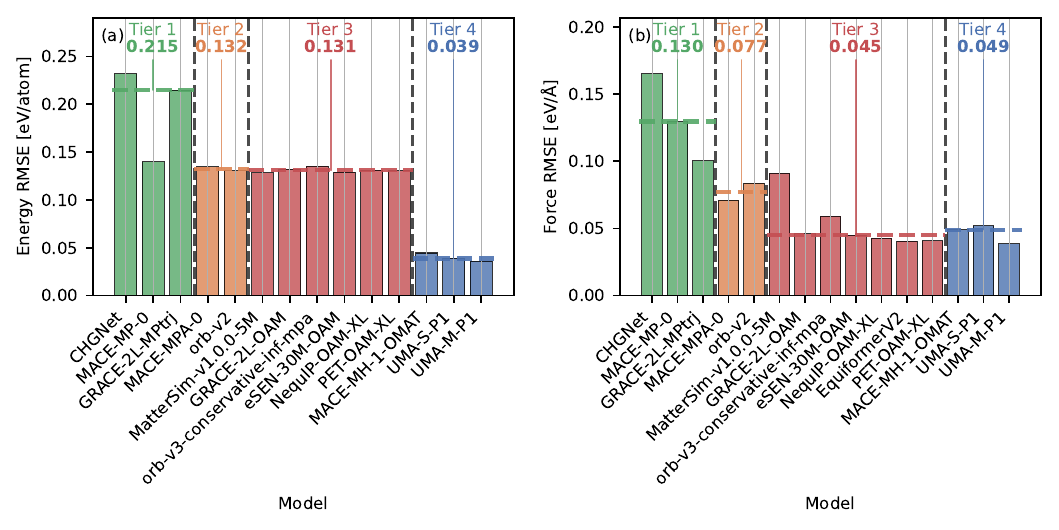}
    \caption{(a) Mean energy RMSE and (b) mean force RMSE for each model averaged over 12 systems, with molecular crystals excluded. Within each tier, the models are shown according to the increasing number of model parameters. Horizontal dashed lines indicate the median mean energy or force RMSE for each model tier.    }
    \label{fig:plot-SI-energy-force-rmses-no-molecular-crystals}
\end{figure}

\newpage

\section*{Pressure histogram error}

\begin{figure}[H]
    \centering
        \includegraphics[width=\linewidth]{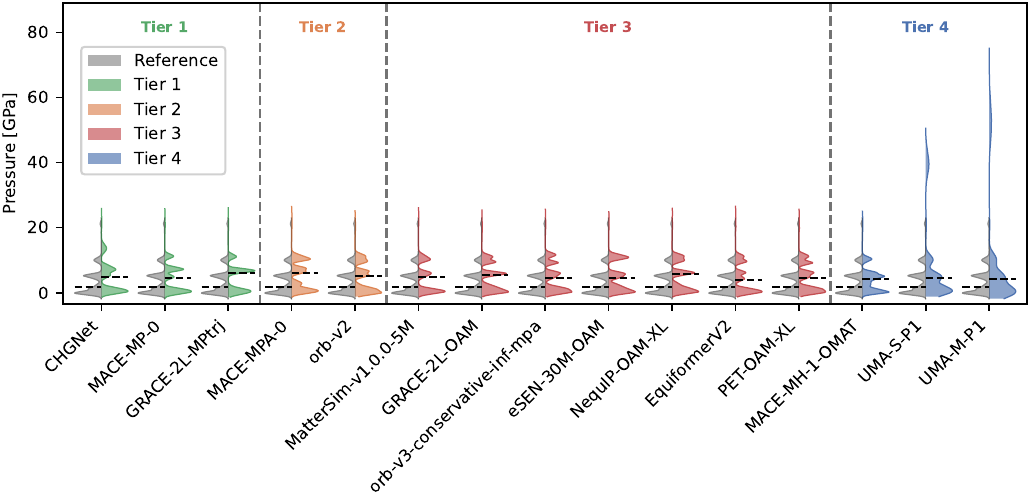}

    \caption{Split violin distributions of MD trajectory pressure values for each model, pooled across the 11 systems with available reference pressure data. Each violin is split vertically: the left half (grey) shows the reference pressure distribution and the right half shows the corresponding model prediction, both estimated via kernel density estimation. White horizontal bars mark the median of each half.}
    \label{fig:plot-SI-pressure-violin-plot}
\end{figure}

\newpage

\begin{figure}[H]
    \centering
        \includegraphics[width=\linewidth]{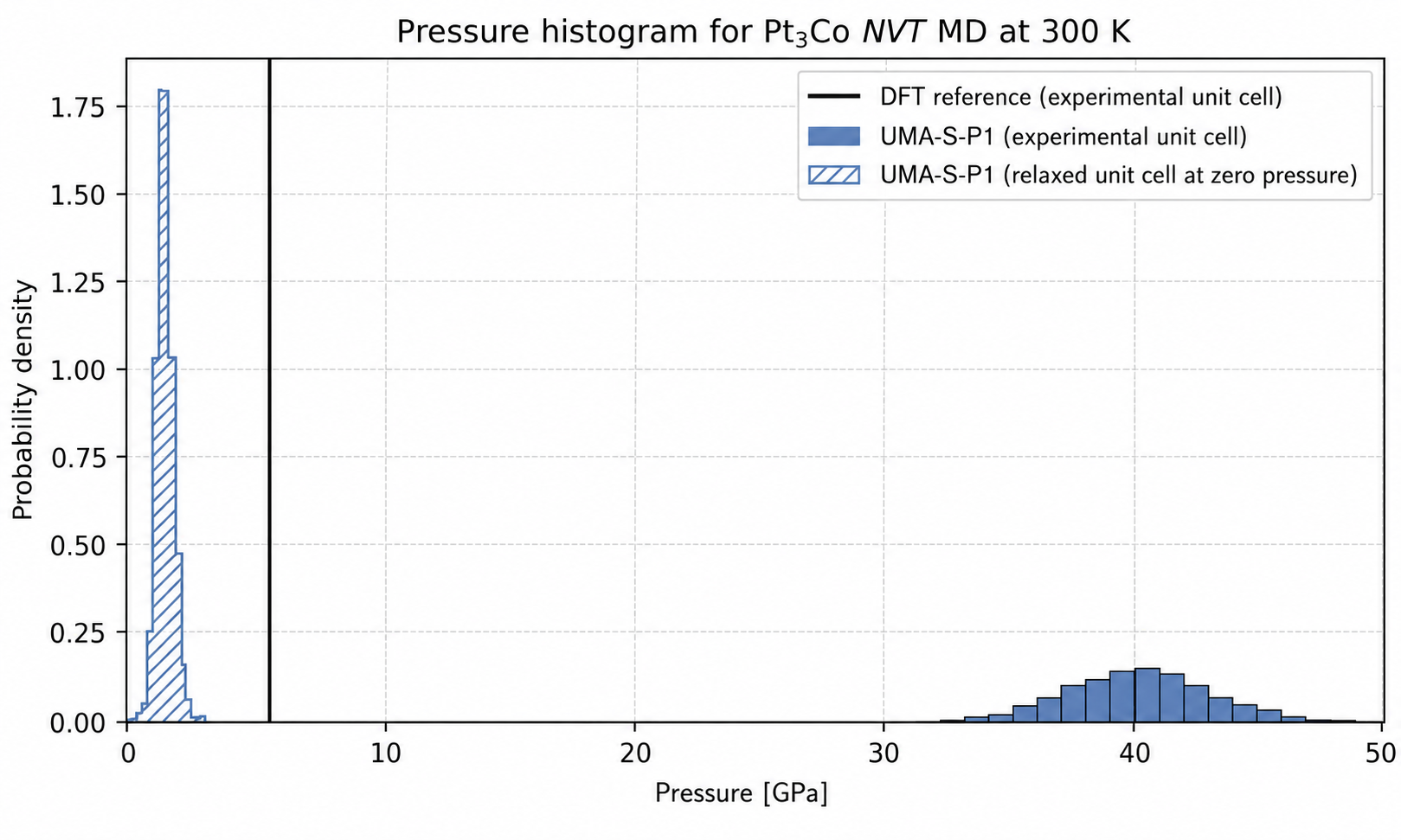}

    \caption{Pressure histogram for Pt$_3$Co $NVT$ molecular dynamics at 300~K. The vertical black line indicates the mean pressure of the DFT reference trajectory generated using the experimental unit cell. The filled blue histogram shows the instantaneous hydrostatic pressure distribution from a fixed-cell UMA-S-P1 trajectory initialised from the same experimental unit cell. The hatched blue histogram shows the corresponding distribution after relaxing the initial structure with UMA-S-P1 before the $NVT$ run. In the relaxed-cell protocol, both atomic positions and lattice vectors were optimised using a Frechet cell filter with the FIRE optimiser, zero external pressure, an $f_{\max}=0.01$~eV~\AA$^{-1}$ convergence threshold, and a maximum of 500 optimisation steps. The subsequent $NVT$ trajectory was initialised with Maxwell-Boltzmann velocities and propagated using a Nose-Hoover chain thermostat at 300~K, with a 1~fs timestep, a 20~fs thermostat damping time, a 10~ps burn-in period, and a 22~ps production trajectory with configurations recorded every step. Histograms are normalised to probability density.}
    \label{fig:plot-SI-pressure-pt3co-uma-s-relaxed-cell}
\end{figure}

\newpage

\begin{figure}[H]
    \centering
        \includegraphics[width=\linewidth]{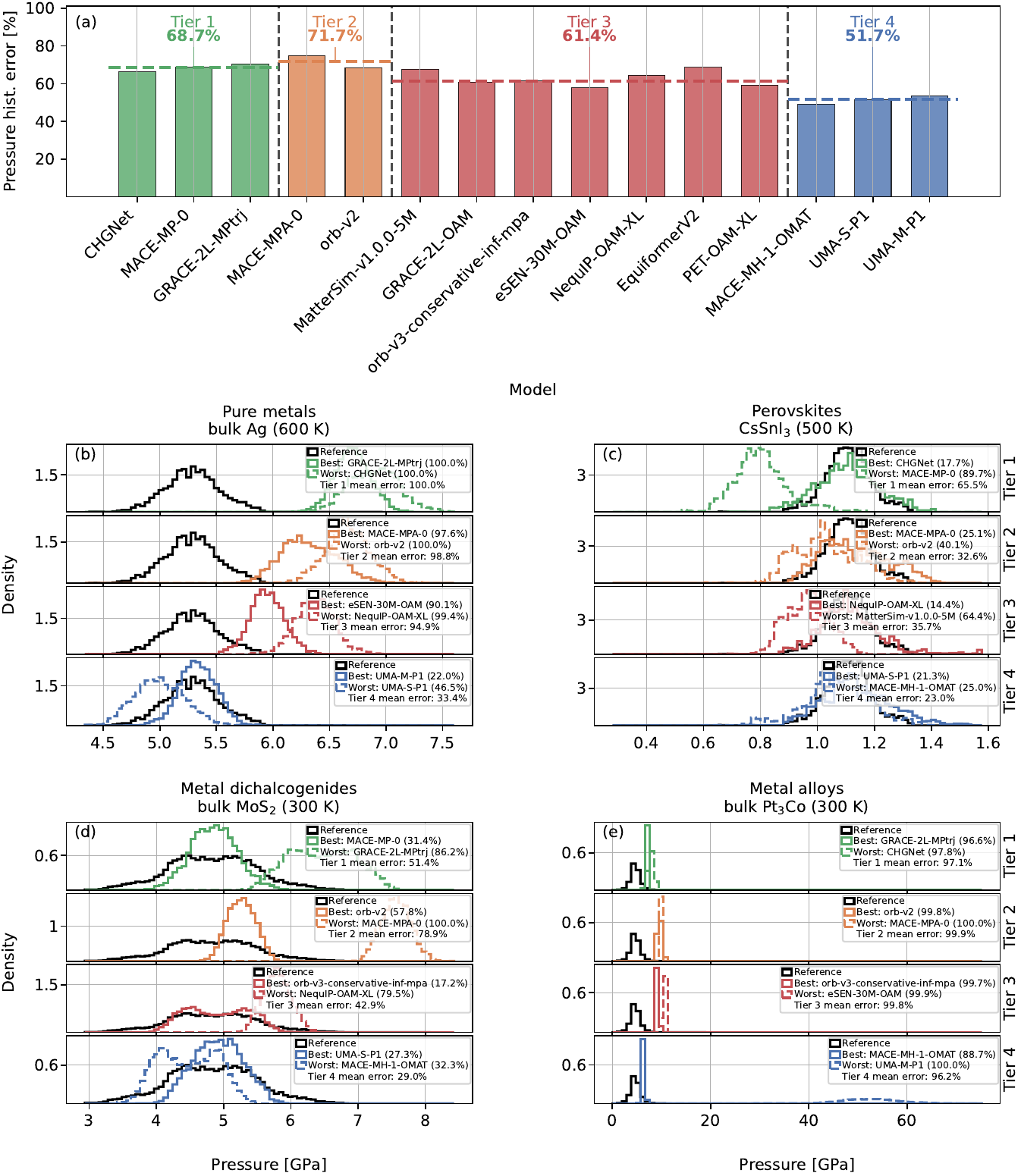}
    \caption{
    Unit-cell pressure distribution panel.
    (a): pressure histogram error for the MLIPs grouped in four tiers. The error is taken with reference to the first-principles molecular-dynamics simulations. Dashed lines indicate the mean pressure error for each model tier.
    (b--e): pressure histograms for the worst-performing system from each material class.}
    \label{fig:plot-SI-pressure-histogram-error-panel}
\end{figure}

\newpage

\section*{Assessment of metrics}

\begin{figure}[H]
    \centering
    \includegraphics[width=\linewidth]{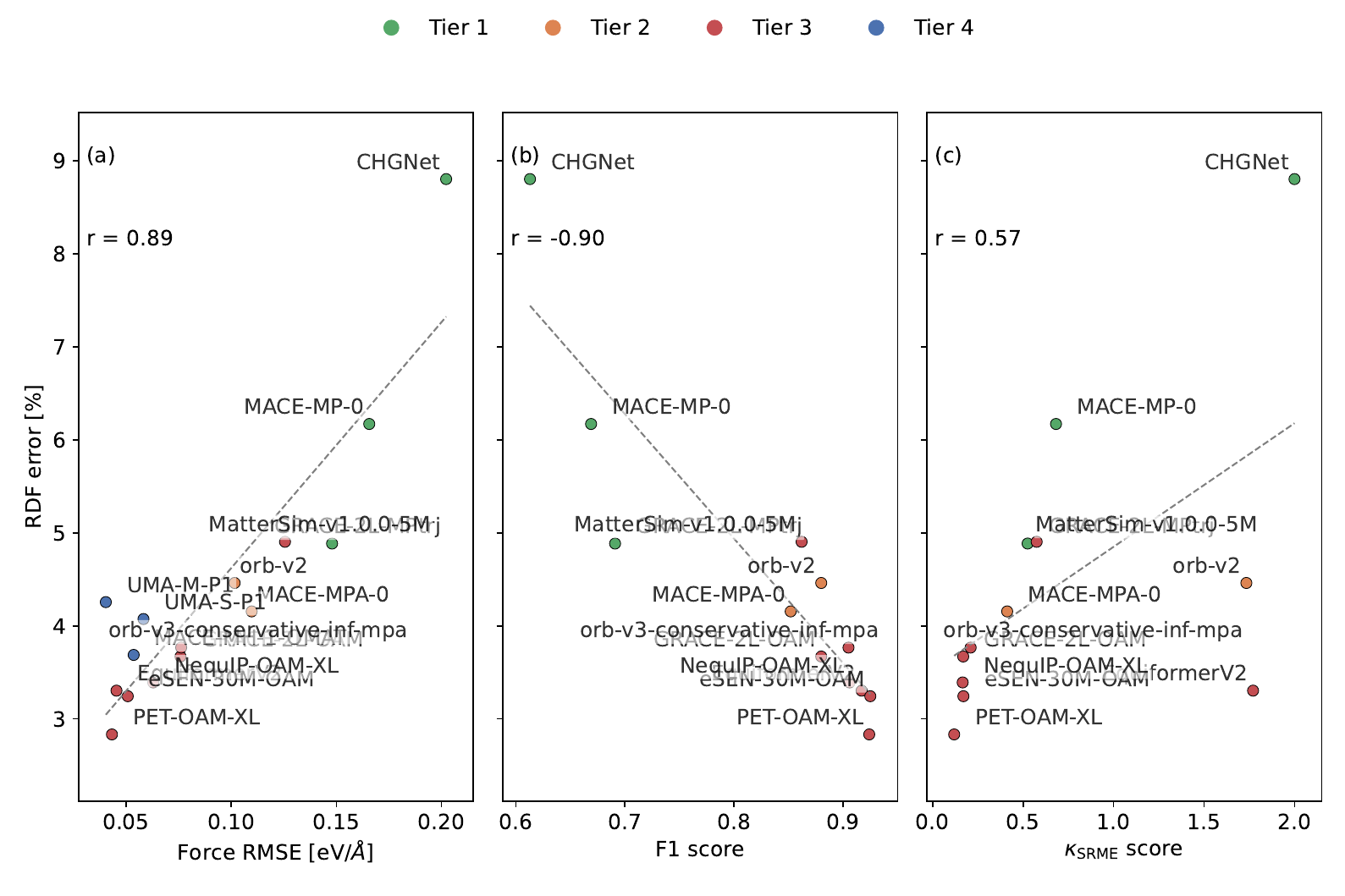}
    \caption{(a--c) RDF correlation panel. In each subplot, the value of the Pearson correlation $r$ is given in the top left corner.
    Each observable error is correlated with respect to the force RMSE, F1 score, and $\kappa_{\text{SRME}}$, respectively. Tier-4 MLIPs are included only in the force RMSE correlation subplot.}
    \label{fig:plot-SI-rdf-correlations}
\end{figure}

\newpage

\begin{figure}[H]
    \centering
    \includegraphics[width=\linewidth]{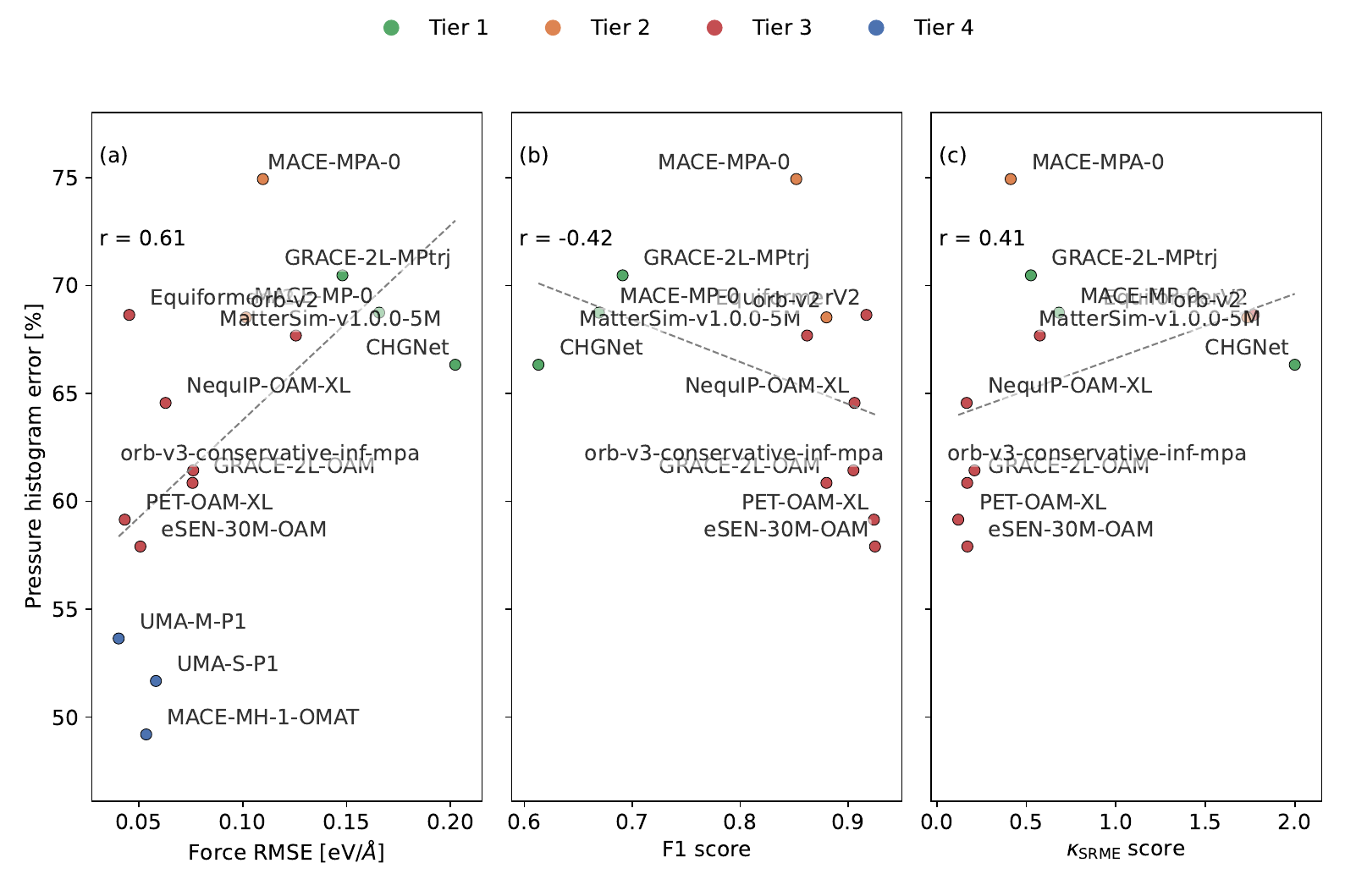}
    \caption{(a--c) Pressure correlation panel. In each subplot, the value of the Pearson correlation $r$ is given in the top left corner.
    Each observable error is correlated with respect to the force RMSE, F1 score, and $\kappa_{\text{SRME}}$, respectively. Tier-4 MLIPs are included only in the force RMSE correlation subplot.}
    \label{fig:plot-SI-pressure-histogram-correlations}
\end{figure}

\newpage

\begin{figure}[H]
    \centering
    \includegraphics[width=\linewidth]{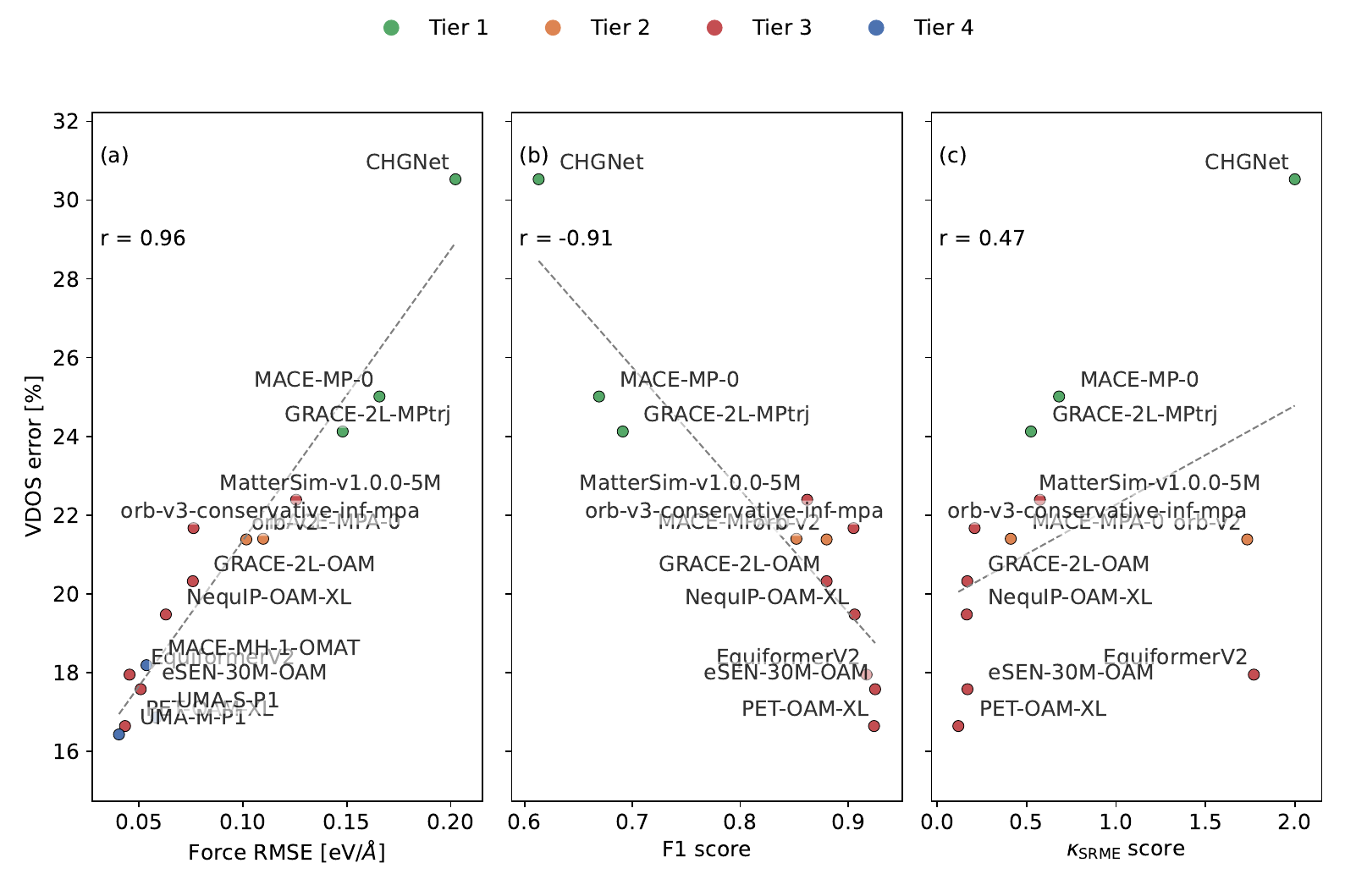}
    \caption{(a--c) VDOS correlation panel. In each subplot, the value of the Pearson correlation $r$ is given in the top left corner.
    Each observable error is correlated with respect to the force RMSE, F1 score, and $\kappa_{\text{SRME}}$, respectively. Tier-4 MLIPs are included only in the force RMSE correlation subplot.}
    \label{fig:plot-SI-vdos-correlations}
\end{figure}

\newpage

\section*{RDF, pressure, VDOS correlations}
\begin{figure}[H]
    \centering
    \includegraphics[width=\linewidth]{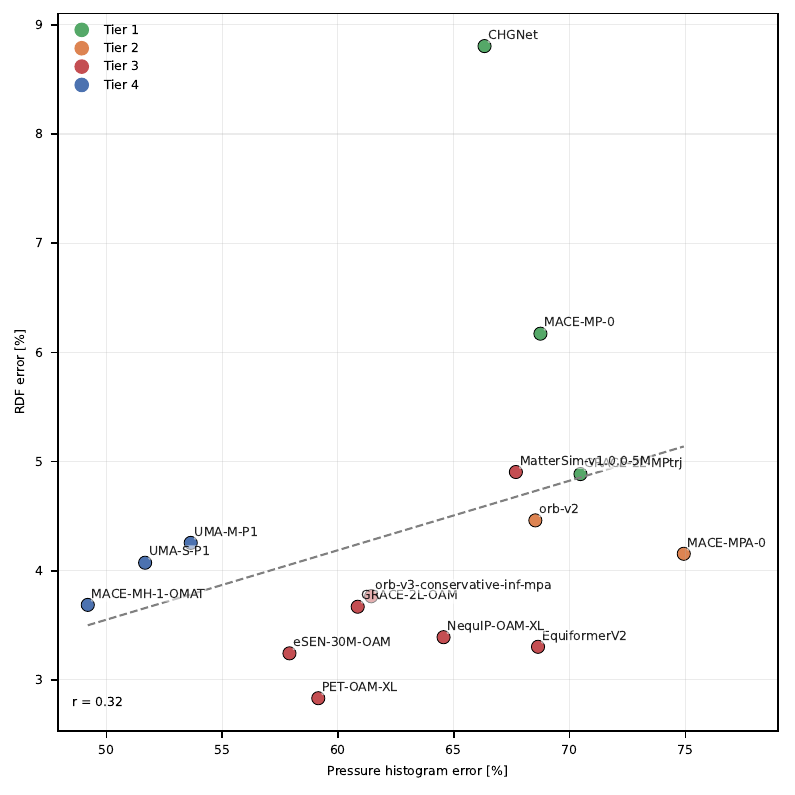}
    \caption{RDF and pressure correlation}
    \label{fig:plot-SI-correlation-rdf-pressure-same-length}
\end{figure}

\newpage

\begin{figure}[H]
    \centering
    \includegraphics[width=\linewidth]{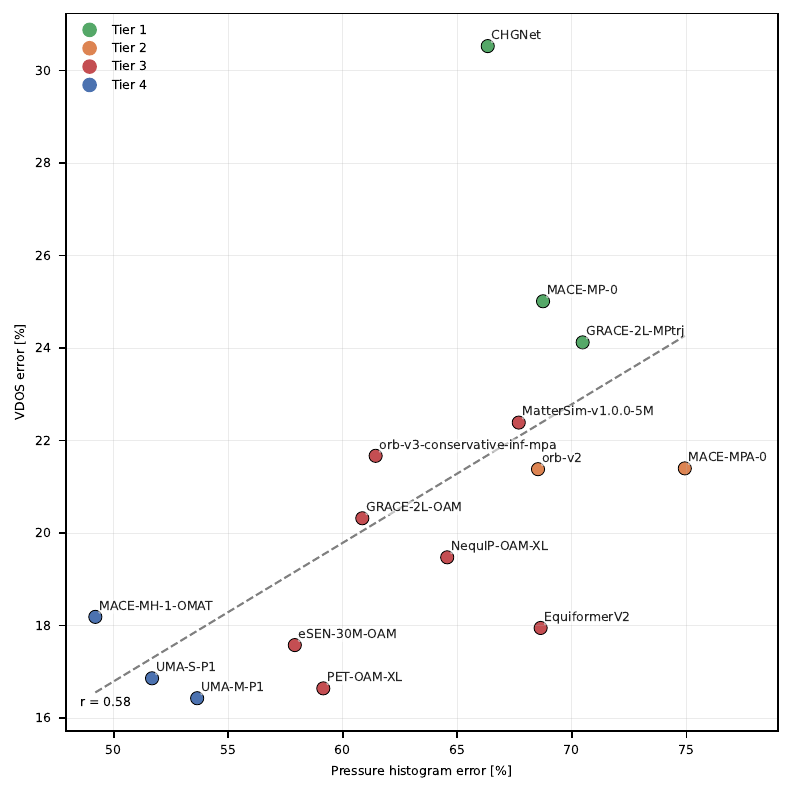}
    \caption{VDOS and pressure correlation}
    \label{fig:plot-SI-correlation-pressure-vdos-same-length}
\end{figure}

\newpage

\begin{figure}[H]
    \centering
    \includegraphics[width=\linewidth]{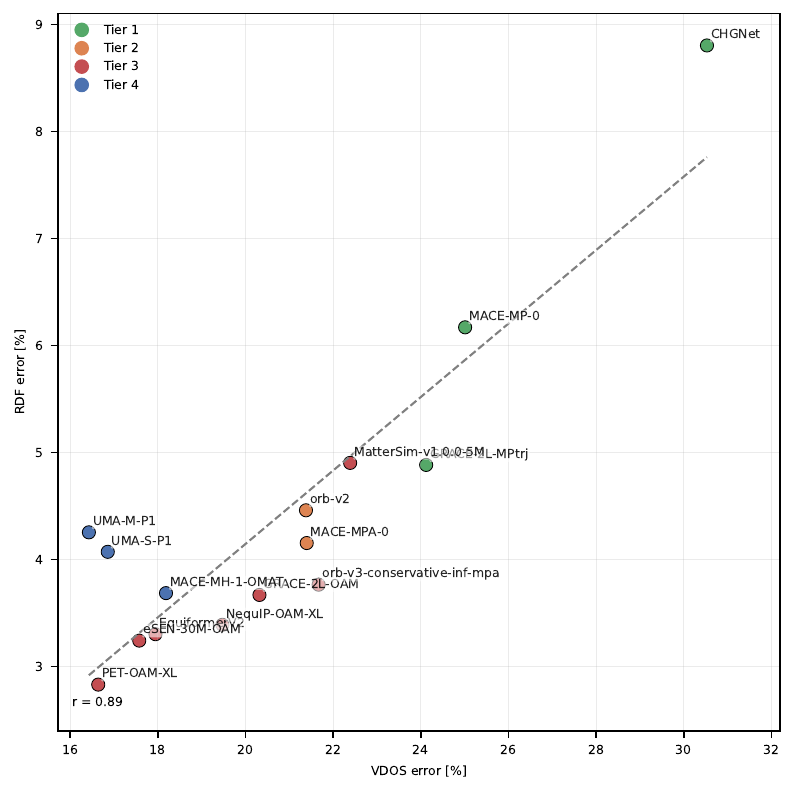}
    \caption{RDF and VDOS correlation}
    \label{fig:plot-SI-correlation-rdf-vdos-same-length}
\end{figure}

\newpage

\section*{Combined similarity metric - Pareto front}

\begin{figure}[H]
    \centering
    \includegraphics[width=\linewidth]{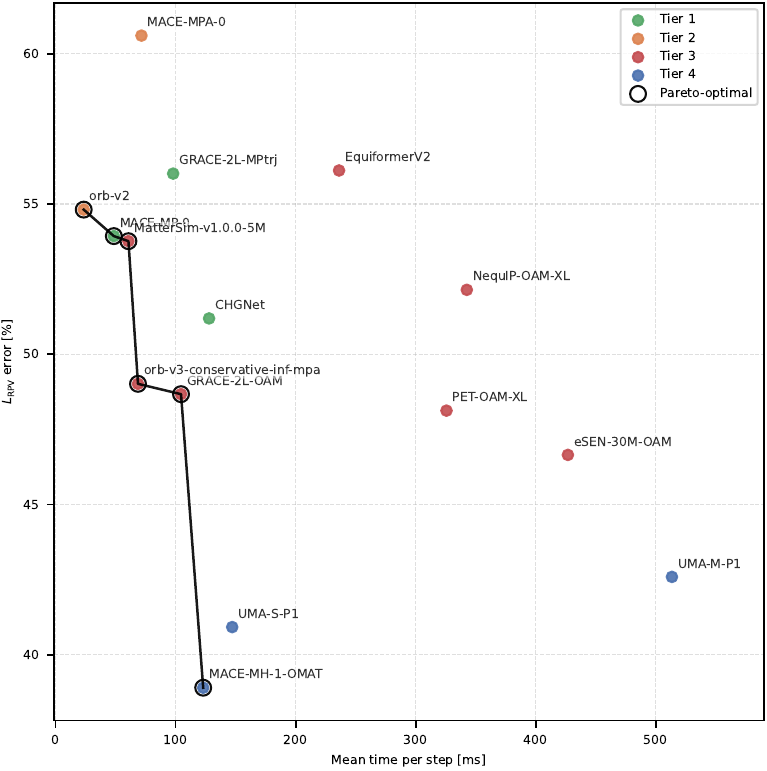}
    \caption{Pareto front of the combined Lehmer-like error metric and the mean time per MD step. The colours denote four model tiers, and the Pareto-optimal MLIPs are indicated with the black circle.}
    \label{fig:plot-pareto-vdos-rdf-panel}
\end{figure}

\newpage

\section*{Observables -- Pareto fronts}
\begin{figure}[H]
    \centering
    \includegraphics[width=\linewidth]{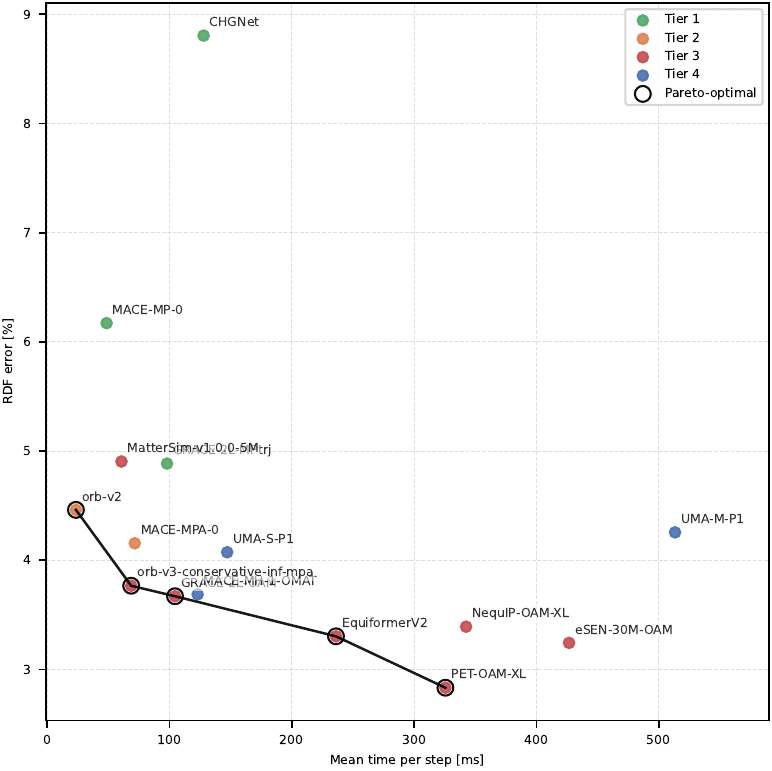}
    \caption{Pareto front -- RDF error, mean time per MD step.}
    \label{fig:plot-SI-pareto-rdf-time}
\end{figure}

\newpage

\begin{figure}[H]
    \centering
    \includegraphics[width=\linewidth]{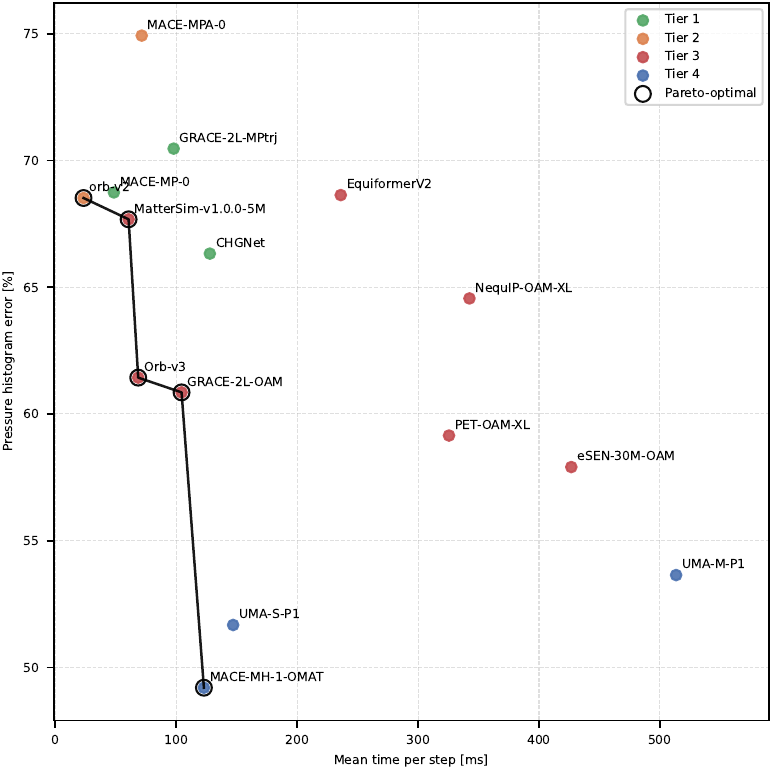}
    \caption{Pareto front -- pressure histogram error, mean time per MD step.}
    \label{fig:plot-SI-pareto-pressure-time}
\end{figure}

\newpage

\begin{figure}[H]
    \centering
    \includegraphics[width=\linewidth]{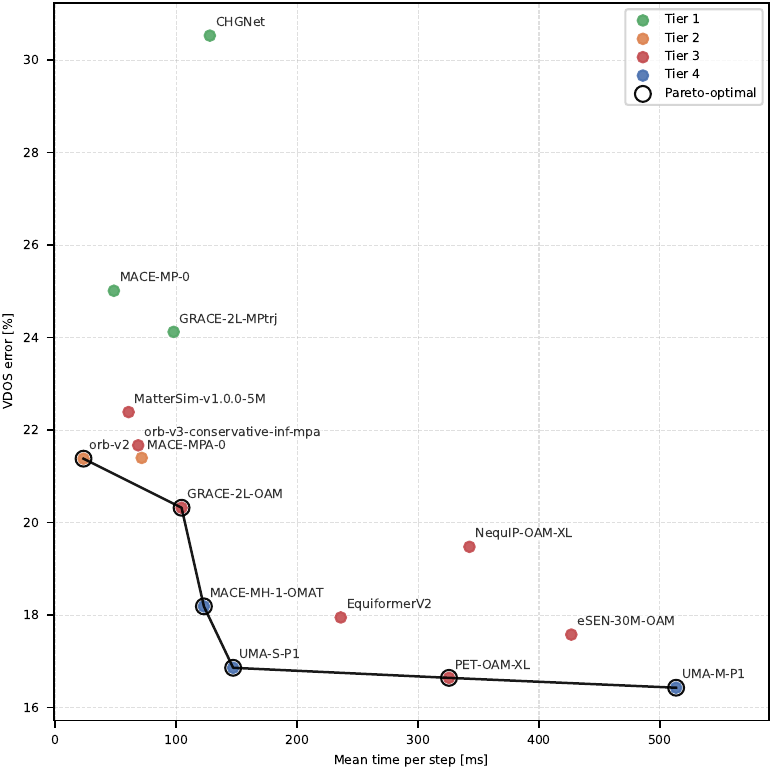}
    \caption{Pareto front -- VDOS error, mean time per MD step.}
    \label{fig:plot-SI-pareto-vdos-time}
\end{figure}

\newpage

\end{document}